\documentclass[conference]{IEEEtran}
\usepackage{graphicx}
\usepackage{color}
\usepackage{amsmath}
\usepackage{amsfonts}
\usepackage{url}
\usepackage[vlined,linesnumbered,algoruled]{algorithm2e}
\usepackage{multirow}
\usepackage{subfigure}
\usepackage{colortbl}
\usepackage{tabularx}
\usepackage{cite}

\definecolor{lightgrey}{rgb}{0.9,0.9,0.9}

\newcolumntype{C}[1]{>{\centering\arraybackslash}p{#1}} 
\newcolumntype{R}[1]{>{\raggedleft\arraybackslash}p{#1}} 

\newcommand{\changefont}[3]{
\fontfamily{#1} \fontseries{#2} \fontshape{#3} \selectfont}

\newcommand{\openbox}{\leavevmode
  \hbox to.4em{%
  \hfil\vrule
  \vbox to.65em{\hrule width.6em\vfil\hrule}%
  \vrule\hfil}}

\providecommand{\qedsymbol}{\openbox}

\newcommand{\argmax}{\operatornamewithlimits{arg\,max}}
\newcommand{\argmin}{\operatornamewithlimits{arg\,min}}

\newcounter{rewritedefcounter}
\newcounter{myDefCounter}

\newcounter{myExampleCounter}
\newenvironment{myExample}
{
	\refstepcounter{myExampleCounter}
	\vspace{0.2cm}
	\newline
	{ \changefont{ptm}{m}{sc} Example \arabic{myExampleCounter}:\ }
	\normalfont
}
{
	\hfill\quad\hbox{\qedsymbol}
	\vspace{0.2cm}
}

\begin{document}
\sloppy

\title{GutenTag: A Multi-Term Caching Optimized Tag Query Processor for Key-Value Based NoSQL Storage Systems}


\author{Christian von der Weth   \and   Anwitaman Datta}

\author{\IEEEauthorblockN{Christian von der Weth, Anwitaman Datta}
\IEEEauthorblockA{Nanyang Technological University (NTU), Singapore\\
Email: \{vonderweth$|$anwitaman\}@ntu.edu.sg}
}

\maketitle

\begin{abstract}
NoSQL systems are more and more deployed as back-end infrastructure
for large-scale distributed online platforms like Google, Amazon or
Facebook. Their applicability results from the fact that most
services of online platforms access the stored data objects via
their primary key. However, NoSQL systems do not efficiently support
services referring more than one data object, e.g. the term-based
search for data objects. To address this issue we propose our
architecture based on an inverted index on top of a NoSQL system.
For queries comprising more than one term, distributed indices yield
a limited performance in large distributed systems. We propose two
extensions to cope with this challenge. Firstly, we store index
entries not only for single term but also for a selected set of term
combinations depending on their popularity derived from a query
history. Secondly, we additionally cache popular keys on gateway
nodes, which are a common concept in real-world systems, acting as
interface for services when accessing data objects in the back end.
Our results show that we can significantly reduces the bandwidth
consumption for processing queries, with an acceptable, marginal
increase in the load of the gateway nodes.
\end{abstract}

\section{Introduction}
\label{sec:Introduction}

A large number of present-day online platforms typically rely on
custom-made distributed NoSQL systems for managing a bulk of their
data -- e.g., Amazon's \textsc{Dynamo}~\cite{DeCandia07Dynamo},
Facebook's \textsc{Cassandra}~\cite{Lakshman09Cassandra}, or
Google's \textsc{BigTable}~\cite{Chang06BigTable} -- instead of
using a full-fledged relational database. NoSQL systems implement a
key to value map as basic data structure, featuring a hash table
like interface to access the data. Their successful application in
online platforms derives from the fact that most relevant queries
can be translated into simple primary-key accesses to the data
store. Examples are to get the tags of web page, the profile of a
user or the features of a product. Compared to traditional solutions
based on relational database systems, the simple data model of NoSQL
systems scales very well in terms of performance, availability,
reliability and maintenance in large-scale distributed settings.

The information needs of online platforms, however, are not fully
limited to key-based queries, i.e. queries that can be translated
into primary key accesses to data objects. From an information
retrieval perspective this issue has been addressed by means of
distributed inverted indexes, mapping individual terms, e.g. tags,
to data objects such as web pages containing that term. Since
multi-term queries represent the majority of user queries, various
approaches utilizing multi-term inverted indexes have been
proposed~\cite{Podnar07Scalable,Chen10TSS}. Here, given a document
with $n$ terms, the number of possible term combinations is in
$O(2^n)$. Thus, restricting to a meaningful subset of multi-term
keys is a crucial design consideration of the proposed systems.
However, these existing works in literature assume static documents,
i.e. the set of terms for a document does not alter. In most online
applications, however, data objects may change over time. With that,
not only the size of the index but also the bandwidth needed to
propagate updates to the index is an issue. Thus, the effect of the
number of index entries on the overall performance is more
pronounced than with static data objects.

In this paper, we investigate the effect of an evolving knowledge base on the application of a distributed inverted index with the support of term combinations. As our first major contribution, we present the results of comprehensive data analysis under the aspect of term combinations.

Firstly, we analyze the tag data from two popular online platforms,
\textsc{Delicious} and \textsc{Flickr}, to quantify the frequency
and distribution of term combinations. From that we can derive the
effect of the support of term combinations on the size and growth of
an inverted index, motivating the requirement to limit the number of
stored term combination. Further, these data sets allow the
estimation for the expected activity of users, i.e., how frequently
users add or delete tags.

Secondly, we measure the frequency and distribution of term
combinations in a real-world query log (\textsc{AOL}). The gained
insights clearly show that the popularity of term combinations
derived from the query history is a meaningful approach for
selecting term combinations to be stored in the inverted index.

Our second contribution is the design and evaluation of a tagging
platform based on a multi-term inverted index. We assume the
widely-used architecture of popular online platforms using a
large-scale, distributed NoSQL system as back end to provide the
services for the overlying applications. While such architectures
inherently scale very well for accesses to the data using the
primary key of data objects, we focus on the efficient support of
queries referring to several data objects at a time. We basically
deploy the concept of an inverted index to map the relevant
characteristics (e.g. set of tags), derived from the information
needs of a service, to the identifier of the data objects. However,
the straightforward application of an inverted index does not scale
in large, distributed systems~\cite{Li03Feasibiltu}. To this end, we
propose and evaluate two extensions to our inverted index
infrastructure:

\textit{(1) Query-driven support of multi-term keys.} Similar to
existing approaches, we store multi-term keys in the inverted index.
However, due to the dynamic characteristics, we note that an a
priori computation of a meaningful number of multi-term keys to be
indexed is not practical. We therefore aim for query-driven (caching
like) optimization techniques, storing only keys that frequently
occur in incoming queries. To efficiently handle changing data
objects, we propose \textit{incremental updates}. Obviously there is
a trade-off regarding the costs for processing queries and
maintaining the index, particularly in the presence of updates, that
we explore.

\textit{(2) Caching of keys on gateway nodes.} We assume that
queries cannot be issued to any arbitrary node in the back end, but
that there exists a smaller set of nodes or resources that act as
the gateway between the application and the back end system. Given
this architecture, we cache a subset of keys on these gateway nodes
to minimize the access to the NoSQL back end. Again, we derive the
set of cached keys from their popularity. Caching the most popular
keys will increase their average load of gateway nodes compared to a
node in the back end. We evaluate the expected increase in the load
depending on the number of back end and gateway nodes.

Next, Section~\ref{sec:RelatedWork} reviews related work to put our
approach in context. Section~\ref{sec:SystemOutline} outlines the
basic architecture of our envisioned tagging platform based on
distributed back end using a NoSQL system.
Section~\ref{sec:DataAnalysis} shows and discusses the result of our
tag data and query log analysis. Section~\ref{sec:queryProcessing}
presents our approach for a query processor on top of a distributed
multi-term inverted index, including a cost analysis and the
discussion of design alternatives. Section~\ref{sec:Index} covers
the index and cache management, particularly the query-driven
identification of popular keys for indexing and caching, and the
handling and propagation of updates in the presence of evolving
data. Section~\ref{sec:Evaluation} features our exhaustive
evaluation, quantifying the effect of multi-term keys, caching and
update frequency on the overall system performance.
Section~\ref{sec:conclusions} concludes.

\section{Related Work}
\label{sec:RelatedWork} This paper contributes to two broad topics:
the analysis of web data and distributed information retrieval on
top of NoSQL systems.
\\
\\
\textbf{Analysis of web data.} Folksonomies or Social Tagging
systems -- allowing users to freely add tags to resources (images,
videos, web pages, etc) -- are currently one of the most popular
ways to organize information on the web. The most noticeable feature
of all folksonomies is that the distribution of tags show power law
relationships~\cite{Gupta10Survey}, i.e. a small subset is popular,
while most other tags occur relatively infrequently. To give some
example, the authors
of~\cite{wetzker2008analyzing,antonellis2009tagging} and
of~\cite{Sigurbjornsson08Flickr} show this for \textsc{Delicious}
and \textsc{Flickr} respectively. Our data analysis of the tagging
data extends these results to tag combinations of various sizes.
In~\cite{Carman09AStatistical} the authors compare the vocabulary
used for tagging and for searching. They find that both vocabularies
are similar. We come to similar conclusions, allowing us to use the
query log of a web search engine to query tagging data, allowing
evaluation of our query processor under realistic workloads, and
obtaining meaningful results for the same.

The analysis of query logs poses a well established means to provide
valuable information in order to improve online
searching~\cite{Silverstein99Analysis,Spink02FromESex,Chau05Analysis,Pass06APicture,Fagni06Boosting,Adar07Why,Baeza08Design,Weber10TheDemographics}.
Regarding basic characteristics of user queries that are relevant in
our context, all works yield similar results. Firstly, the average
query contains 2-4 terms and more than 2/3 of all queries contain
more than one search term. Additionally, comparing the results from
different years clearly shows slow but continuous increase of these
figures. This motivates our support term combinations as keys within
an inverted index. Secondly, both the frequency of queries and of
query terms show a power law relationship. Thus, a few queries are
very frequent, while majority of queries occur only once or a few
times. In our analysis we show that this holds also for term
combinations of various sizes derived from search queries. This
motivates our approach for a query-driven identification of term
combinations to be stored in the inverted index.
\\
\\
\textbf{NoSQL systems and information retrieval.}
\textsc{Dynamo}~\cite{DeCandia07Dynamo},
\textsc{Cassandra}~\cite{Lakshman09Cassandra},
\textsc{Voldemort}~\cite{Voldemort} and
\textsc{BigTable}~\cite{Chang06BigTable} are some well known
distributed NoSQL implementations used in the back end of some
popular online platforms. The need for NoSQL systems is driven by
the requirements of large-scale online platforms (performance,
availability, scalability). In their core, all NoSQL systems
implement a key to value map featuring a hash table like
\texttt{put/get/delete} interface to insert, access and update the
data. They mainly differ in their expressiveness with respect to
processing queries, their support of (semi-)structured data and
application-specific characteristics. The successful application of
key-value stores arises from the fact that most services provided by
online platforms access the required date using their primary key.
However, various important services, particularly the term-based
search for data objects, are not efficiently supported in key-value
stores.

There are two principal approaches to support queries that are not based on the primary keys of data objects:
(1) Divide \& Conquer approaches, like the \textsc{MapReduce} framework~\cite{Dean08MapReduce} (used in, e.g., \textsc{BigTable}), essentially `ignore' the underlying key to value map. Here, the initiating node sends a query to all nodes in the network. Each node, then, evaluates the query on its locally stored data, and sends the result back to initiating node. Finally, this node combines all partial results to the final result. Since contacting the nodes and locally processing a query is done in parallel, the response time is good. However, due to the involvement of each node for each query, the induced overhead in terms of resources and bandwidth is high. Thus, such an approach is more suitable for batch and pre-processing.
(2) Distributed inverted indexes map individual terms, e.g. tags, to documents, e.g. web pages, containing that term in order to facilitate information retrieval. More and more NoSQL implementations natively support inverted indexes (e.g., \textsc{BigTable}, or \textsc{Cassandra}). This makes their maintenance transparent for programmers but prohibits the realization of customized optimizations. Without tailored implementations of inverted indexes, i.e. adding, deleting and updating index information on the application level, programmers cannot consider application-specific characteristics to optimize such additional indexes.

Multi-term queries -- which represents the majority of user queries
-- are evaluated by merging the corresponding list of documents for
each query term~\cite{Reynolds03Efficient,Tang04Hybrid}. Although
optimization techniques to reduce the bandwidth consumption, like
\textit{Bloom Filter}~\cite{Reynolds03Efficient} exits, the costs
for multi-term searches using single-term inverted indexes are
generally very high~\cite{Li03Feasibiltu}. As a result, various
approaches utilizing multi-term inverted indexes have been
proposed~\cite{Skobeltsyn07WebText,Chen08EfficientMultiKeyword,Chen10TSS}.
Given a document with $n$ terms, the number of possible term
combinations is in $O(2^n)$. Thus, the limitation to a meaningful
subset of term combinations is a crucial part of the proposed
systems. However, all the existing approaches assume static
documents, i.e. the set of terms for a document does not alter.
However, in Web 2.0 applications, the set of tags of a resource
changes over time.

The presented work, \textsc{GutenTag}, specifically focuses on and
takes into account such dynamics of the workload. For that, not only
the size of the index but also the bandwidth needed to propagate
updates to the index needs to be taken into account. To make the \textsc{GutenTag} back end scale, we designed
novel mechanisms for indexing and processing multi-term queries efficiently, even in the presence of frequent updates
We believe these mechanisms are of more general interest, benefiting keyword-based search techniques in similar NoSQL-powered platforms or decentralized environments deploying Distributed Hash Tables (DHT).

The results from query log analyses, particularly the power law
distribution of query and query terms, strongly indicates the need
of caching mechanisms. Components caching the results for popular
queries are integral part of existing search engines. From an
academic perspective the issue of result caching has been addressed
by various works,
e.g.~\cite{Baeza07TheImpact,Gan09Improved,Baeza08Design,Fagni06Boosting}.
These works differ in the strategies they propose to identify and
update the set of cached query results. With our focus on the
effects of a multi-term inverted index on the system performance, we
propose a rather simple caching scheme on top of the index. However,
our results show that we can nevertheless significantly boost the
performance through caching, which adds to the benefits of the
multi-term indexing approach.

\section{System Outline}
\label{sec:SystemOutline}
\begin{figure}
	\centering
		\includegraphics[width=0.45\textwidth]{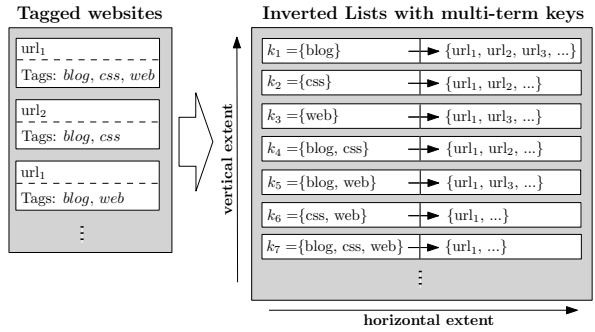}
	\caption{Indexing scheme}
	\label{fig:indexing_scheme}
\end{figure}

As a typical application scenario for NoSQL storage systems, consider online platforms that allow users to tag resources, i.e. to add or delete tags over time. To give examples, resources can be images or video clips of media-sharing sites (e.g., \textsc{YouTube}, \textsc{Flickr}), products or services of online sales or auction sites (e.g., \textsc{eBay}, \textsc{Amazon}), or websites for social bookmarking (e.g., \textsc{Delicious}). Resources are identified by their unique url. Similar to many recent online applications we assume principally a custom-made distributed key-value store for managing all resources. The urls of resources represent their keys; the resources and all relevant information about them, including the tags, represent the values in the storage system.
\\
\\
\textbf{Distributed inverted index.}
Identifying resources using their key reflects the frequent task of accessing single resources directly, i.e. to get all information about an image, product or website. We aim to efficiently support keyword-based queries to address sets of resources that all are tagged with set of terms of the user query. To accomplish this, we deploy the concept of a distributed inverted index. Since multi-term queries represent the majority of user queries (see Section~\ref{subsec:QueryLogAnalysis}), we favour a multi-term inverted index. Here, the keys for the key-value store derive from the combinations of terms/tags; values are the urls of the pages tagged with the corresponding term combinations. Figure~\ref{fig:indexing_scheme} illustrates the approach.

\begin{figure}
	\centering
		\includegraphics[width=0.4\textwidth]{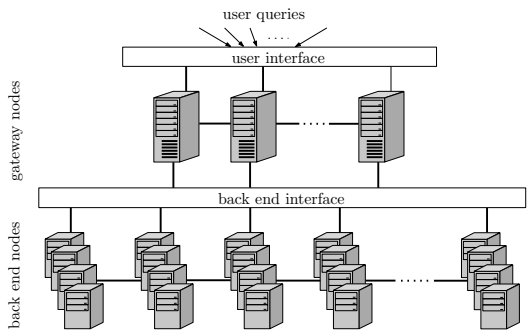}
	\caption{Sytem architecture}
	\label{fig:architecture}
\end{figure}

We distinguish between \textit{single-term keys}, i.e. keys derived from one single term and \textit{multi-term keys}, i.e. keys derived from a combination of terms. In principle, the number of possible multi-term keys to identify a resource grows exponentially with the number of its tags. Since the set of tags associated with a resource may change over time, not only the size of the index but particularly the bandwidth needed to propagate updates to the index is also an issue. To keep only a limited but meaningful set of multi-term keys, we propose a query-driven selection of keys. In a nutshell, we store only the keys derived from popular term combinations, i.e. from term combinations that frequently occur in the recent history of past user queries. Thus, an important issue to address is the expected trade-off between the benefit of multi-term keys in order to improve the query processing performance and the induced additional overhead in the presence of updates.
\\
\\
\textbf{System architecture.}
Emulating existing online platforms, we consider a hybrid architecture -- though simplified compared to real-world architectures -- comprising of a distributed back end for storing the application data, and dedicated components for coordinating task (e.g., access control, monitoring, etc.). See Figure~\ref{fig:architecture}. Throughout the paper, we distinguish between gateway nodes back end nodes.

\textit{Gateway nodes.}
In real-world systems, services typically do not access the data by contacting an arbitrary node in the back end. Instead, services requests are sent to a limited set of dedicated resources -- henceforth called gateway nodes -- that access the back end to retrieve the data and send them back to the requester. We exploit this fact to cache keys, i.e., storing a meaningful set of keys from the inverted index on the gateway nodes to minimize the bandwidth-consuming access to the distributed back end. We distribute the cache over the gateway nodes using a Distributed Hash Table (DHT), i.e., each gateway node is responsible for a specific range of keys of the inverted index. All nodes have complete routing information, so a node can forward a key/lookup to the correct node in a single hop.
Once a node receives a query $q$ it forwards $q$ to the node for the key derived from $q$. Thus, a repeat query is always handled by the same gateway node. This node then accesses its local cache and the back end to answer $q$ and eventually returns the result.

\textit{Back end nodes.} The back end comprises of up to several hundred, typically rather low-cost, nodes. All back end nodes are organized in a Distributed Hash Table (DHT). We assume that each back end node maintains enough routing information to allow for a $O(1)$ routing, i.e., the access to a single key requires only a constant and very low number of hops. The back end nodes serve as underlying infrastructure for the distributed inverted index. We further assume that all single-term keys are available in the index and therefore each query can be answered. (Flexibly and efficiently re-inserting popular single-term keys requires additional mechanisms and is beyond the scope of this article.) Each back end node evaluates the popularity of its keys derived from its access history. Using this local statistics a back end node (a) retrieves and stores the inverted list of popular multi-term keys or (b) forwards the inverted list of popular single-term and multi-term keys to the cache, i.e. to the corresponding gateway node responsible for a key.

\section{Data Analysis}
\label{sec:DataAnalysis}
We use publicly available tagging and query log data sets. In both cases, we focus on the distribution of term combination, either derived from the set of tags of a resource or from a search query. While the results are interesting in themself, they also specifically affect the design of our distributed inverted index and query processor. The results of the tagging data analyses let us derive practical values for important parameters of the inverted index and highlight the necessity to identify a meaningful subset of term combinations to index; the results of the query log analysis show how to identify such a set based on the popularity of term combinations.

\subsection{Tagging data}
\label{subsec:taggingdata}
\begin{figure}
	\centering
		\includegraphics[width=0.48\textwidth]{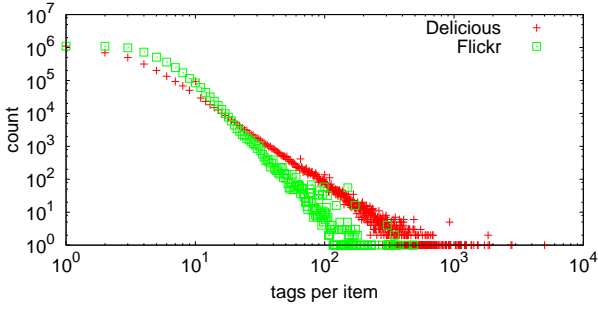}
	\caption{Number of tags per resource}
	\label{fig:both_tag_distribution}
\end{figure}

We use the datasets from two very popular platforms \textsc{Delicious} and \textsc{Flickr}. As a social bookmarking site, users of \textsc{Delicious} tag bookmarks to websites and share them online. In \textsc{Flickr}, being a media sharing site, users can upload and tag photos and video clips. Both datasets were obtained in 2006. Table~\ref{tab:BasicNumbersOfDatasets} shows the number of resources, the number of distinct tags and the average number of tags per resource for both datasets.
\begin{table}
	\centering
		\footnotesize
		\begin{tabular}{|c|c|c|}
			\hline
																					& \textsc{Delicious} 			& \textsc{Flickr} \\
			\hline\hline
			\#resources  												& 3,441,885 							& 5,626,921				 \\
			\hline
			\#tags (distinct) 									& 981,387									& 812,409					 \\
			\hline
			tags per resource ($\emptyset$) 	& 4.19										& 4.01						\\
			\hline
			actions per minute ($\emptyset$) 	& 66.74										& 107.25						\\
			\hline
		\end{tabular}
	\caption{Basic numbers of datasets}
	\label{tab:BasicNumbersOfDatasets}
\end{table}
The distribution of the number of tags per pages is shown in Figure~\ref{fig:both_tag_distribution}. Not unexpectedly, the number of tags and corresponding frequency shows basically a power law relationship in both datasets. In direct comparison, \textsc{Flickr} features more resources with a small number of tags ($1-10$), but less resources with a large number of tags compared to \textsc{Delicious}. Further, \textsc{Delicious} features a significant number of resources with a very large number of tags ($> 1000$).
\\
\\
\textbf{Storage Requirements}
The overall required storage for the indexing scheme is mainly determined by the number of list entries. In principle, given a set of tags $T_p$ for a web page $p\in P$, $P$ being the set of all web pages, each possible combination of a subset of those tags are conceivable to form an inverted list. Therefore the number of possible, non-empty list entries for $T_p$ is one less than the size of its power set $|\mathcal{P}(T_p)|-1$ (discarding the empty set). The complete number of non-empty subsets, i.e. number of tuples of tags, is $2^{|T_p|}-1$. With that $S_{total}$ as the required storage for all possible list entries of all pages can be computed as $S_{total}=|P|\cdot \sum_{p\in P} (2^{|T_p|}-1) $. We can re-write the formula using the binomial coefficient to explicitly reflect the various sizes of the possible subsets. To give an example, $\tbinom {|T_p|} {3}$ is the number of all possible subset of size 3 for a give set $T_p$. Using the binomial coefficient results in the following formula:
\begin{equation}
	S_{total}=K\cdot \sum_{p\in P} \left [ \sum_{i=1}^{|T_p|} \binom {|T_p|} {i} \right ]
\label{eq:total_storage_binom}
\end{equation}
An exponential upper bound for the required storage obviously does not scale. However, there are also reasonable means to limit this worst-case behavior, already somewhat indicated by Formula~\ref{eq:total_storage_binom}.
(1) Although the average number of tags per resource is reasonably small (slightly above 4 for both datasets), there are still various resources with a very large number of tags, potentially resulting in a vast number of list entries. However, for meaningfully describing or searching a resource typically a rather small number of tags are sufficient. We therefore limit the set of tags to derive the tag combinations; let $t_{max}$ be the maximum number of considered tags. Further, $C_p$ denotes the subset of $T_p$ containing all tags of page $p$ that are considered for creating inverted lists. Note that for estimating the number of resulting list entries the actual method of how to derive $C_p$ -- e.g. highly rated tags or tags that have been provided by a large number of users -- in case of $|T_p|<t_{max}$ is not relevant.
(2) Several works, e.g.~\cite{Baeza08Design,Spink02FromESex}, and our own query log analysis show that the average number of query terms is between 2 and 3 (later in Section~\ref{subsec:QueryLogAnalysis}). Thus, storing large keys, i.e. keys for large tag sets, is not meaningful since those keys would very rarely by queried. We therefore limit the maximum size of keys, denoted by $s_{max}$. For example, if $s_{max}=4$ we derive only pairs, triplets and quadruplets of tags as keys. Formula~\ref{eq:total_storage_binom_ext} incorporates $C_p$ and $s_{max}$:
\begin{equation}
	S_{total}=K\cdot \sum_{p\in P} \left [ \sum_{i=1}^{\min(|T_p|, t_{max}, s_{max})} \binom {\min(|T_p|, t_{max})} {i} \right ]
\label{eq:total_storage_binom_ext}
\end{equation}
The minimum function ensures the $k\leq n $ requirement for the binomial coefficient $\tbinom {n} {k}$ to be valid. Given $s_{max}$ and $t_{max}$ the upper bound for the number of resulting list entries for a URL is in $O({t_{max}}^{s_{max}})$ reducing the behavior from exponential to polynomial; additionally, in practice the values for both $t_{max}$ and $s_{max}$ tend to be rather small.

\begin{figure*}
\parbox{.30\linewidth}{
	\centering
		\includegraphics[width=0.31\textwidth]{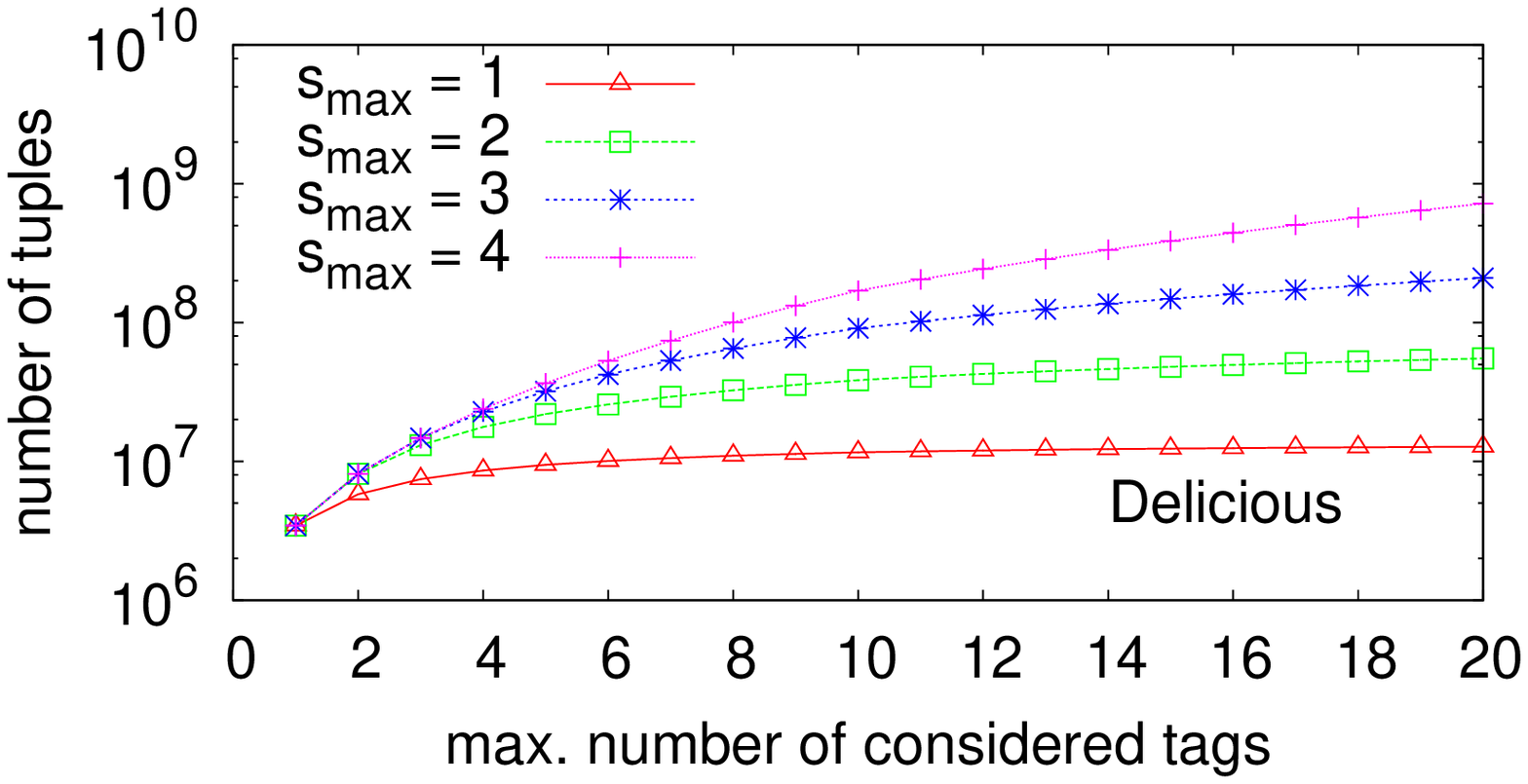}
		\includegraphics[width=0.31\textwidth]{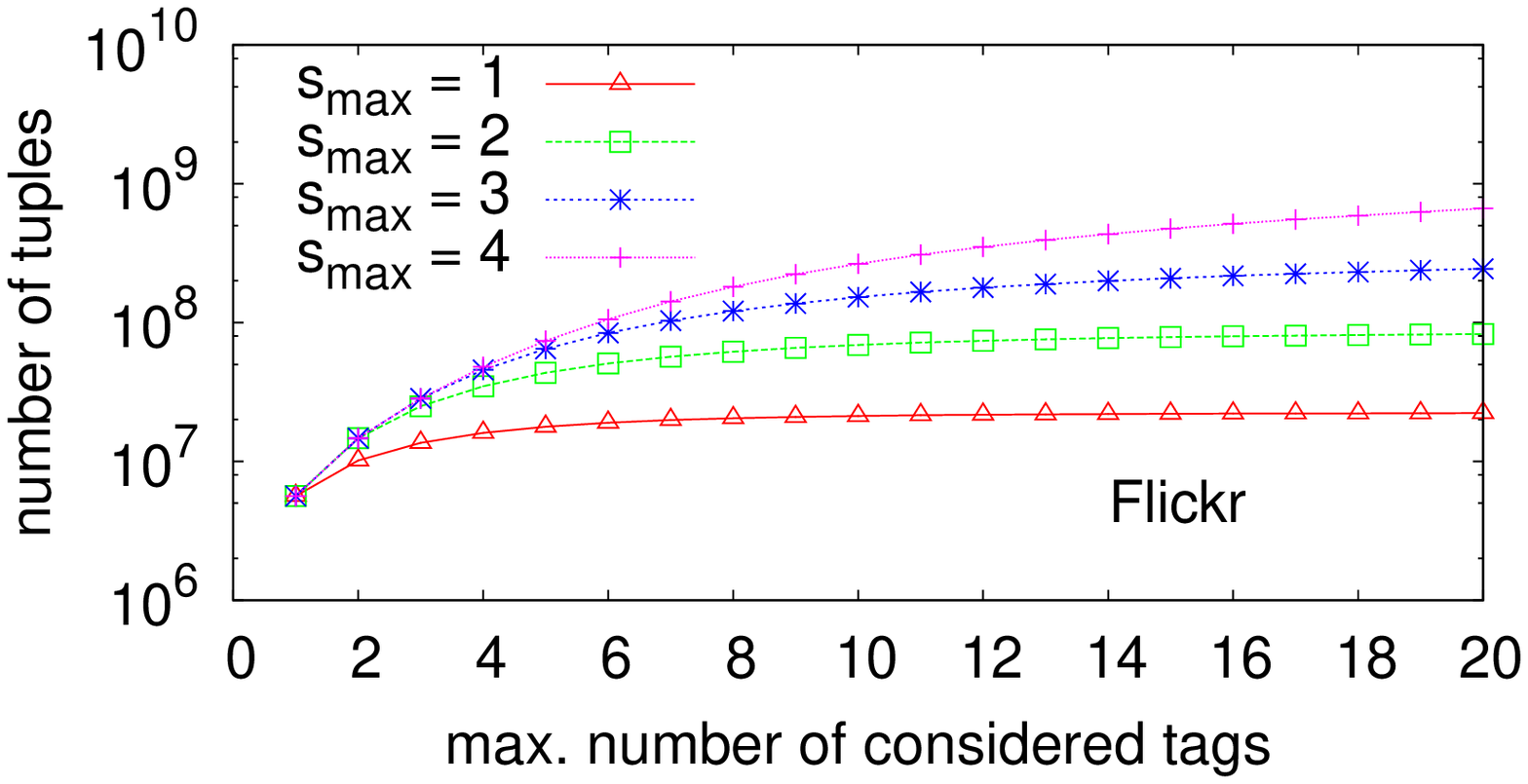}
	\caption{Number of inverted list entries}
	\label{fig:delicious_storage_requirements}
}
\hfill
\parbox{.30\linewidth}{
	\centering
		\includegraphics[width=0.31\textwidth]{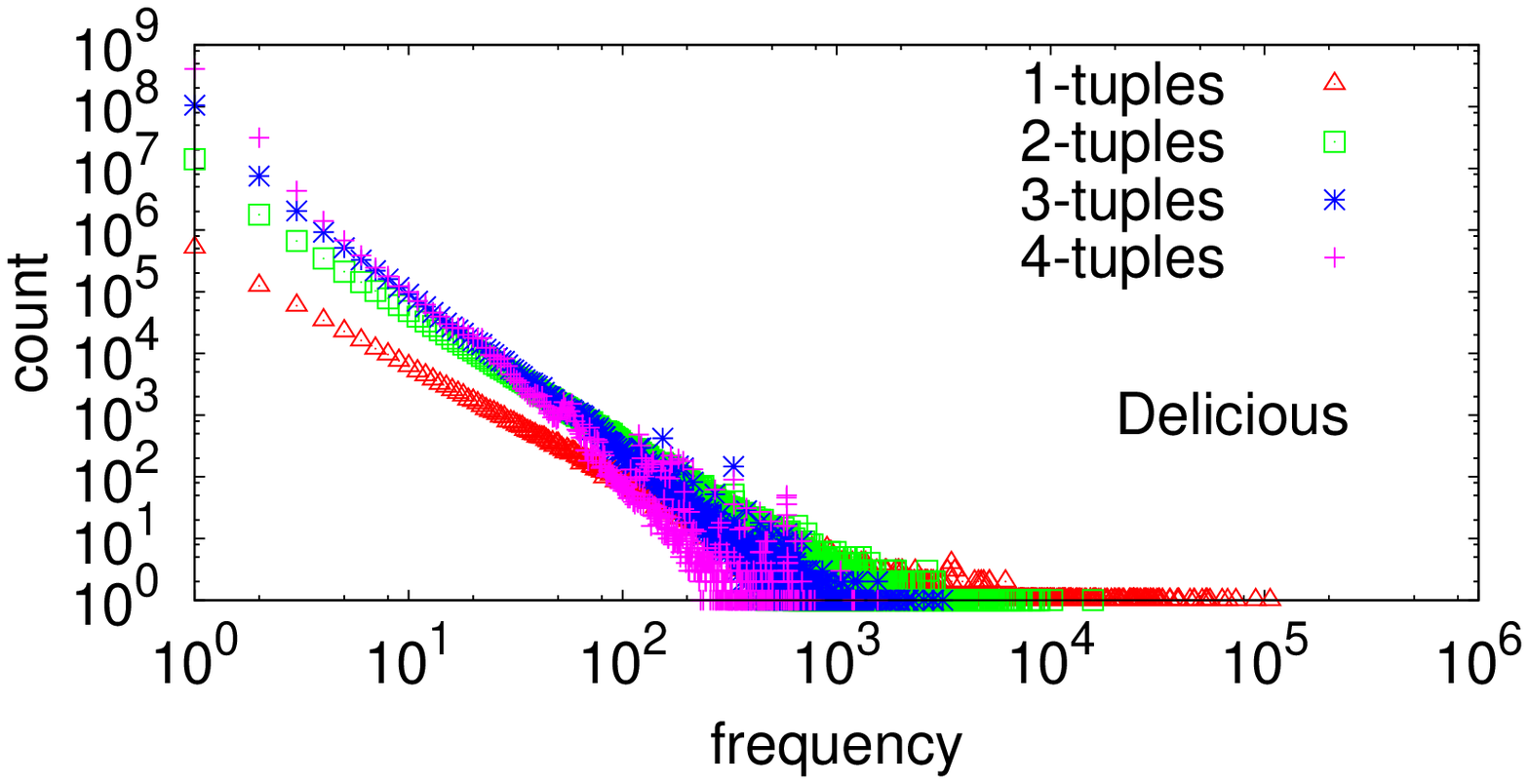}
		\includegraphics[width=0.31\textwidth]{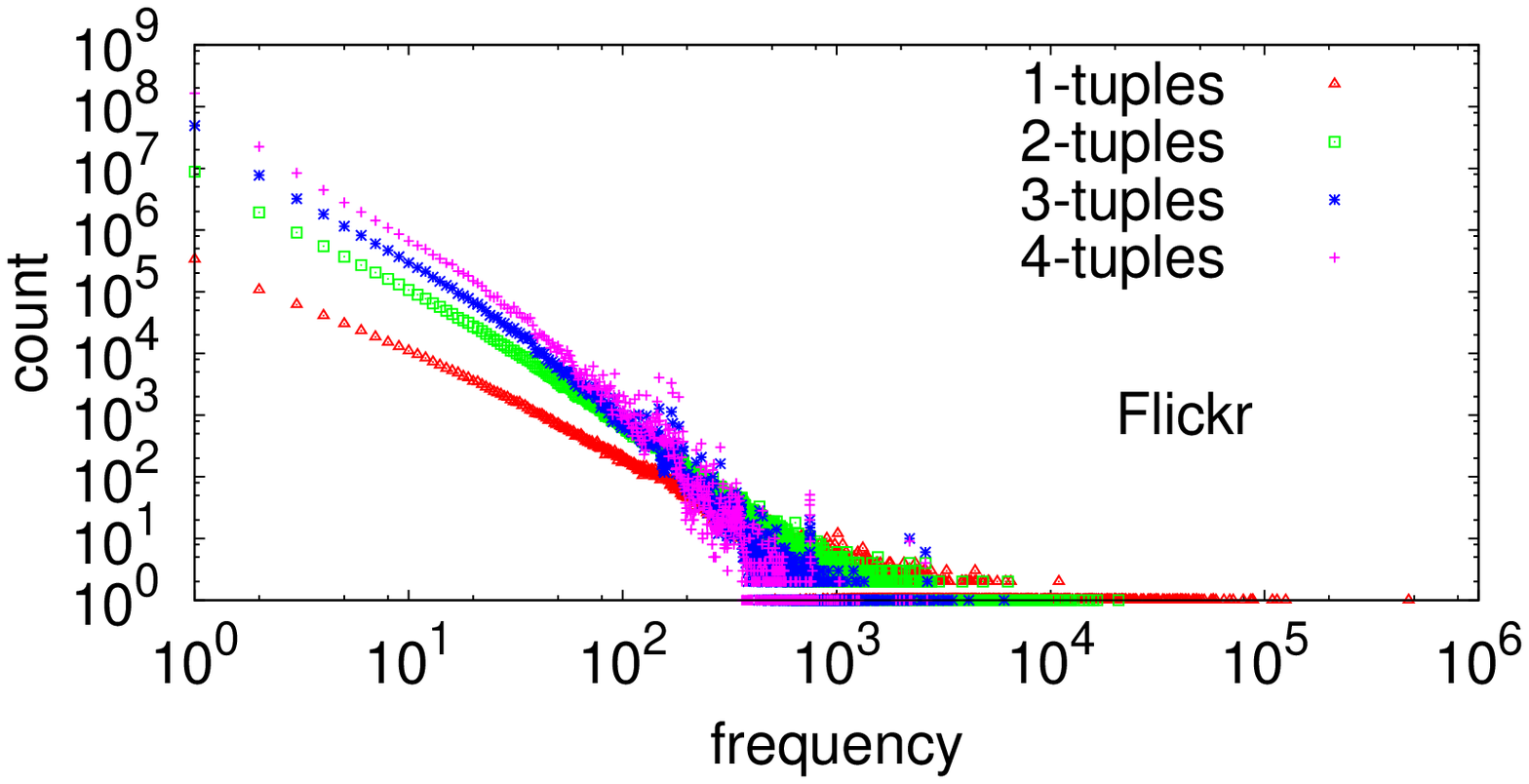}
	\caption{Frequency distribution of tag sets}
	\label{fig:both_delicious_all}
}
\hfill
\parbox{.30\linewidth}{
	\centering
		\includegraphics[width=0.31\textwidth]{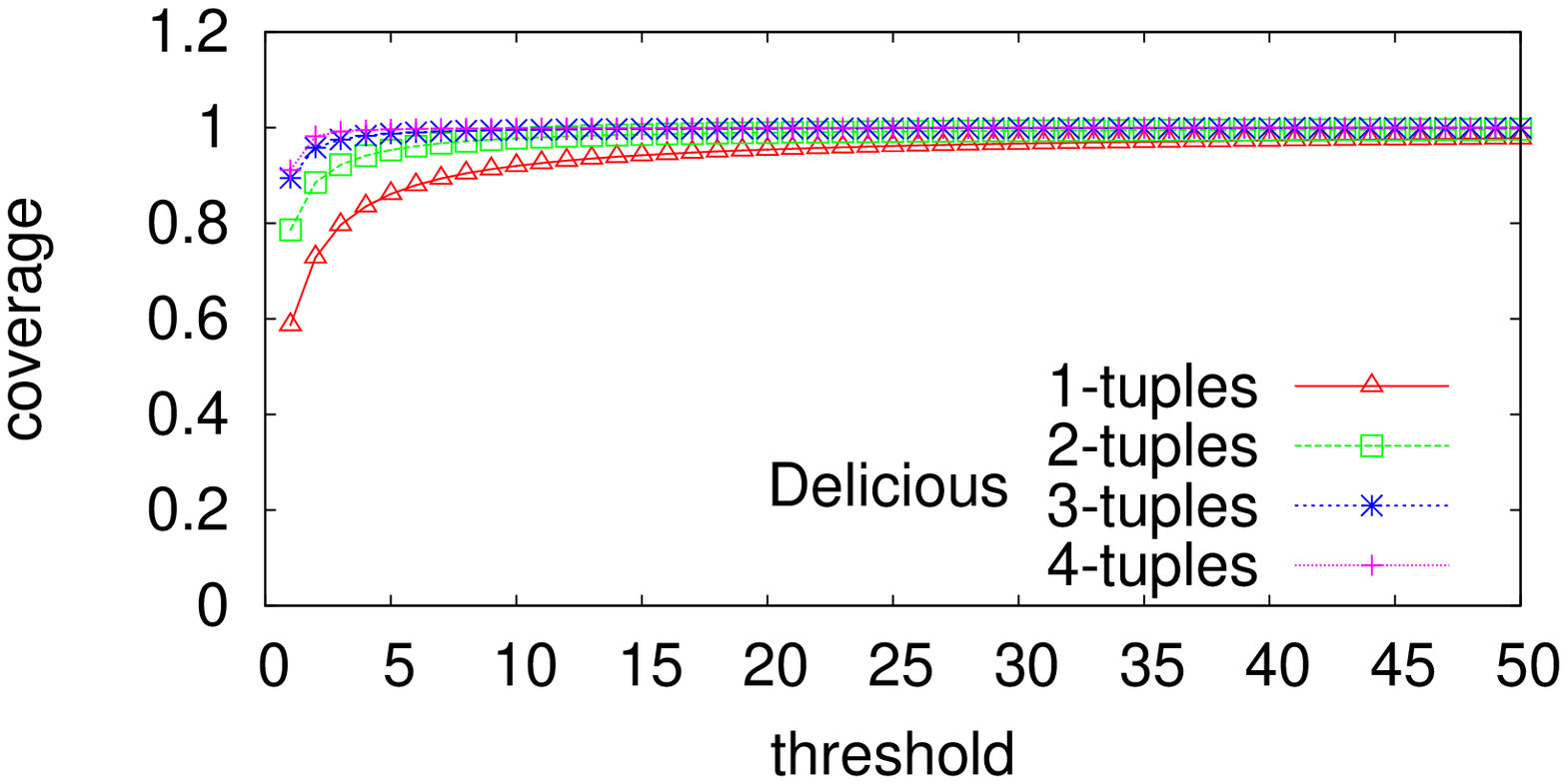}
		\includegraphics[width=0.31\textwidth]{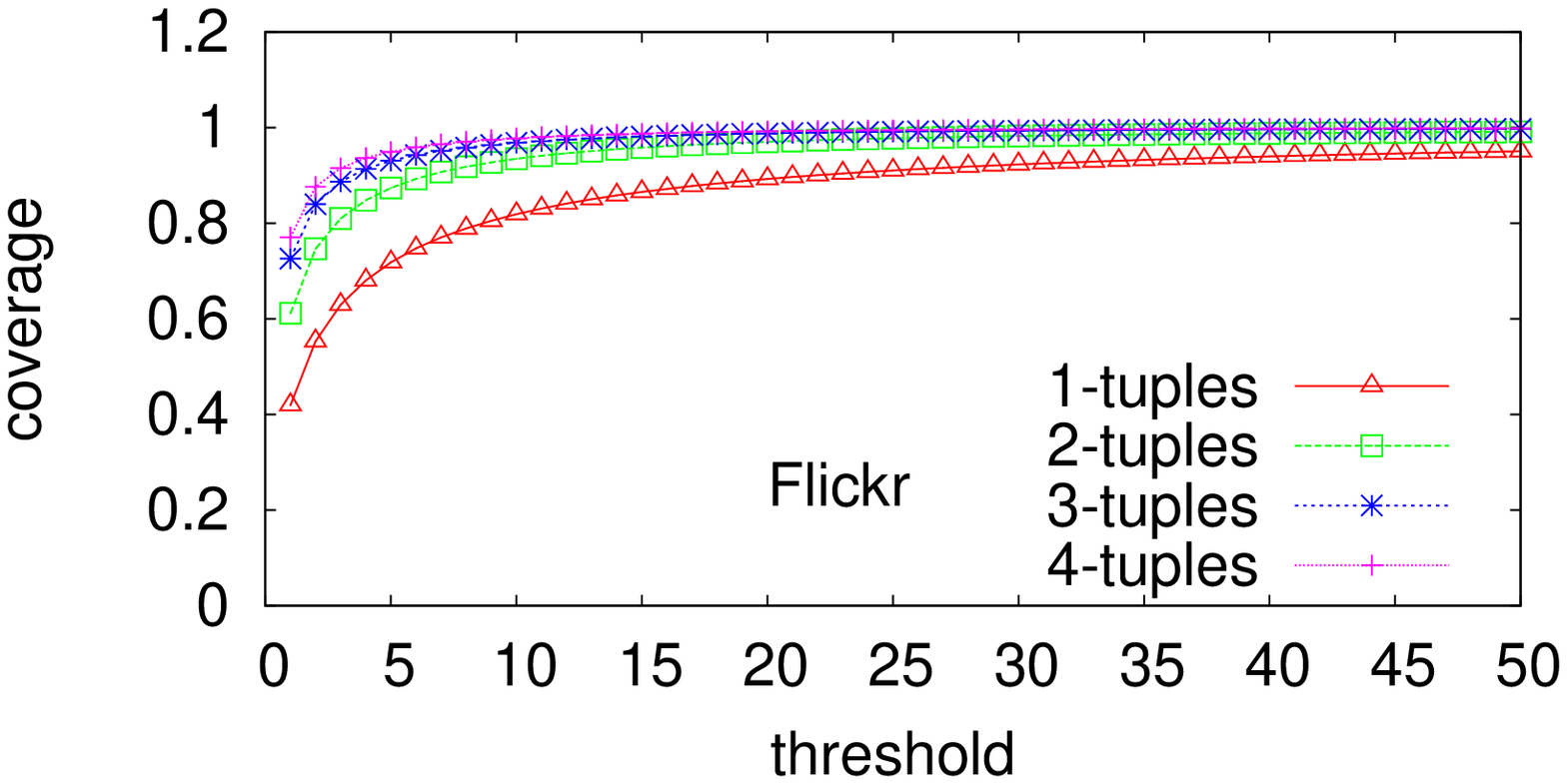}
	\caption{Coverage of complete inverted lists}
	\label{fig:coverage_both_all}
}
\end{figure*}

To quantify these theoretical findings, we computed the number of list entries, for both the \textsc{Delicious} and \textsc{Flickr} dataset, with $1 \leq s_{max} \leq 4$ and $1 \leq t_{max} \leq 20$. Note that for a given $s_{max}$ we also considered all keys $k_i$ of a smaller size, i.e. $1 \leq |k_i| \leq s_{max}$. Figure~\ref{fig:delicious_storage_requirements} shows the results. The qualitative development of the storage requirements for various values for $t_{max}$ and $t_{max}$ are very similar for both datasets; the storage requirements significantly increase if the maximum number of tags per key increase. This clearly indicates that storing all possible keys in the inverted index is not reasonable. Each curve for a value for $s_{max}$ eventually converges to a fixed value. This point is reached, if $t_{max} \geq \argmax_{p \in P}|T_p|$, i.e. all tags of all resources are considered. For example, for $t_{max} = 20$ the percentage of resources where all tags are considered for keys are 98.2\% for \textsc{Delicious} and 99.5\% for \textsc{Flickr}. Regarding the quantitative results both dataset show significant differences. Due to the larger number of resources and distinct tags in the data set, \textsc{Flickr} requires more storage for small values of $s_{max}$. For larger values of $s_{max}$ \textit{Delicious} becomes more storage-consuming, since there, more resources have a larger number of tags yielding more keys per resources (cf.~Figure~\ref{fig:both_tag_distribution}).

Additionally, to give some illustrative numbers Table~\ref{tab:StorageExampleSizes} shows the estimated size of the inverted storage for $t_{max} = 20$. We assume that the entries of an inverted list are URLs pointing to the resources tagged with the corresponding key, i.e. set of tags. According to~\cite{Miranda09Trends} the average URL length is approximately 73 characters.
\begin{table}
	\centering
		\footnotesize
		\begin{tabular}{|l|r|r|r|r|}
			\hline
											& \multicolumn{4}{c|}{$\mathbf{s_{max}}$} \\
			\hline
																& 1						& 2 				& 3 				& 4   \\
			\hline\hline
			\textsc{Delicious}				& 0.8 GB			& 3.7 GB 		& 14.2	GB	& 48.9 GB		\\
			\hline
			\textsc{Flickr} 					& 1.5	GB			& 5.6	GB		&	16.5 GB		&	45.1 GB		\\
			\hline
		\end{tabular}
	\caption{Required storage in GB for $\mathrm{t_{max}=20}$}
	\label{tab:StorageExampleSizes}
\end{table}
\\
\\
\textbf{Horizontal and Vertical Extent}
\label{subsec:HandVextent}
The results for the required storage hide the information about the vertical extent, i.e. the number of distinct keys, and horizontal extent, i.e. the length of the inverted lists, of the index. However, particularly in distributed systems, this information is relevant to estimate the distribution of keys among peers and the average workload of peers. One extreme case is that all pages feature the same set of tags $T = T_r$ for all $r \in R$. Here, the inverted list for each key contains all resources and the number of keys is $\sum_{i=1}^{\min (|T|,s_{max},t_{max})}{\binom{\min(|T|,t_{max}}{i}}$. In the second extreme case, each resource features a unique set of tags $T_r$, i.e. $\bigcap_{r \in R}{T_r}=\emptyset$. In this case, the inverted list for each key contains only one resource and number of keys is $\sum_{r \in R}{ \left[   \sum_{i=1}^{\min (|T_r|,s_{max},t_{max})}{\binom{\min(|T_r|,t_{max}}{i}} \right ] }$. Obviously, the reality is somewhere in between these two extreme, depending on the actual dataset, mainly the distribution of tags. In the following, we therefore analyze the vertical and horizontal extent of the inverted index for the \textsc{Delicious} and \textsc{Flickr} datasets.
\\
\\
\textit{Length of inverted lists.} The length of an inverted list for a key $k$ is specified by the number of resources where $k$ can be derived from the available set of tags. To evaluate this, we computed the frequency of each key among all resources for both datasets. Figure~\ref{fig:both_delicious_all} shows the resulting relationship between the frequency, i.e. length of the corresponding inverted list, and the number of keys with a specific length.
Basically, both datasets yield a power law relationship between the list length and the number of lists with the same length. This is in line with previous observations for single tags~\cite{Sigurbjornsson08Flickr,wetzker2008analyzing,antonellis2009tagging}; here, we also show similar relationships for sets of tags. To better quantify the differences between the power law relationships for different key sizes, $1 \leq |k| \leq 4$ and $t_{max}=20$, we computed the fitting function $f(l) = \alpha \cdot l^\beta$ to extract the scaling factor $\alpha$ and skew $\beta$ depending on list length $l$. Table~\ref{tab:PowerLawComparison} lists the resulting parameter values. The results support our expectation that with keys with larger size the power law relationship shifts more and more to inverted lists of shorter length. However, even for $|k|=4$ there are still some keys with a considerable length for their list.
\begin{table}
	\centering
	\footnotesize
		\begin{tabular}{|l|c||c|c|c|c|}
			\hline
					\multicolumn{2}{|c||}{\multirow{2}{*}{$\alpha \cdot l^\beta$}} 				& \multicolumn{4}{c|}{size of key} \\
			\cline{3-6}
					\multicolumn{2}{|c||}{}								& 1						& 2 				& 3 				 & 4   \\
			\hline\hline
			\multirow{2}{*}{\textsc{Delicious}}		&	$\alpha$	& $5.1\cdot 10^5$			& $1.7\cdot 10^7$ 		&  $1.1\cdot 10^8$	& $4.1\cdot 10^8$	\\
			\cline{2-6}
						&	$\beta$ & $2.0$			& $2.9$ 		& $3.7$	& $3.7$		\\
			\hline \hline
			\multirow{2}{*}{\textsc{Flickr}}		&	$\alpha$	& $3.4\cdot 10^5$			& $8.8\cdot 10^6$ 		& $4.9\cdot 10^7$	& $1.6\cdot 10^8$		\\
			\cline{2-6}
						&	$\beta$ & $1.6$			& $2.1$ 		& $2.6$	& $2.8$		\\
			\hline
		\end{tabular}
	\caption{Power law parameters for $\mathrm{t_{max}=20}$}
	\label{tab:PowerLawComparison}
\end{table}

\begin{figure}
	\centering
		\includegraphics[width=0.48\textwidth]{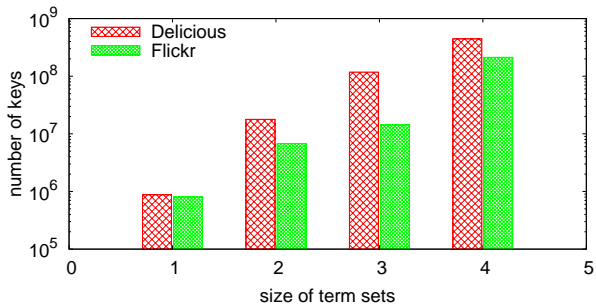}
	\caption{Number of keys}
	\label{fig:flickr_delicious_all}
\end{figure}

In principle, a possible approach to limit the maximum storage size of the index is to limit the length of the inverted list of keys by means of threshold $l_{max}$ specifying the maximum number of entries per list. The rationale is that users typically only view the first top-$k$ results for query. To evaluate the effect of $l_{max}$ on the index, we re-plotted the results to show the percentage of keys with a list of a length $\leq l_{max}$; see Figure~\ref{fig:coverage_both_all}. Particular for keys $k$ with $k > 1$ and a reasonable choice for $l_{max}$, e.g. $l_{max} > 30$, by far most key lists have a length smaller then $l_{max}$. Thus, the storage that can be saved by limiting the maximum length for key lists is rather limited and does  not justify the involved risk of a reduced recall, particularly in case of queries with a number of terms larger than $s_{max}$. An approach to minimize the overall size of the inverted index must therefore mainly focus on reducing of the number of key lists stored in the index.
\\
\\
\textit{Number of keys.} Figure~\ref{fig:flickr_delicious_all} shows the absolute number of keys for various key sizes and for both datasets. In all cases, the number of considered tags $t_{max}$ was set to 20. For the chosen values $1 \leq s_{max} \leq 4$ the number of distinct keys significantly increase the size of the keys. This is due to the fact that average of number of tags per resource is $> 4$ for both datasets. The more the key size exceeds the average number of tags per resource the less resources feature enough tags to derive keys of large size. Thus, the number of distinct keys decreases again for increasing key sizes above the average number of tags (not shown here). However, $s_{max} = 4$ is already a quite large value for practical purposes.

\subsection{Query log analysis}
\label{subsec:QueryLogAnalysis}
While tag datasets can be acquired by crawling existing platforms, acquiring query logs is challenging. Due to privacy concerns, service providers do not make their query logs public\footnote{We contacted both \textsc{Delicious} and \textsc{Flickr} and asked for anonymized query log data. These requests have been declined.}. Further, synthetically created query histories are, in general, inapt to pattern the frequency, popularity, etc. of queries in real-world systems over time.
We therefore use the AOL query log~\cite{Pass06APicture} which is, to the best of our knowledge, the only real-world query log of reasonable size, containing mainly English queries.  The log contains over 28.8 Mio. query requests issued by over 650,000 users and was collected in the period from March to May, 2006. As a convenient coincident, the query log as well as both tagging datasets from \textsc{Delicious} and \textsc{Flickr} have be collected in quite the same period, i.e. in early months of 2006. Naturally, the tagging datasets also include older tags starting from 2003 for \textsc{Delicious} and 2004 for \textsc{Flickr} representing the years when both platforms were officially be launched. For example, in the \textsc{Flickr} datasets, annual details describing the shooting date of a photo belong to the most popular tags. And also in the \textsc{query log} such annual details are part of many queries. Thus, a non-matching query log might lead to distorted evaluation results. Particularly a much more recent log might contain queries referring to events that are not potentially covered by tags.
\\
\\
\textbf{Data pre-processing and cleaning.}
Our data cleaning process for the AOL query log consisted of several steps.
(1) We removed all stop words from the set of terms for each query; to do this we used the Perl module \textsc{Lingua}\footnote{\url{http://search.cpan.org/~creamyg/Lingua-StopWords-0.09/}}.
(2) We removed all queries featuring an URL as the only term. This is true for approximately 25.1\% of all queries, showing clearly that many user ``misuse'' search engines as the browsers address bar.
(3) We removed all terms from queries containing only non-alphanumeric characters. Of course, we kept terms inherently containing non-alphanumeric characters, e.g. ``web2.0'', ``jack's'' (like in many restaurant, diner or pub names) or ``bed\&breakfast''.
(4) We removed all terms from queries with more than 30 characters. Most of these long terms are just gibberish character strings, but also often concatenations of several words not separated by a blank.
(5) We removed all queries that -- after potentially having removed some of their terms -- consisted of more than 100 characters (with around 6,400 queries a marginal number). These queries are often more or less complete sentences, e.g. song lyrics or error messages.
(6) We removed all queries whose complete set of terms were removed in previous processing steps. This was true for about 1.1 Million queries which represents 5.1\% of queries that were still in the query log after the previous cleaning steps. Thus, we only consider non-empty queries with our analysis.

\begin{figure*}[htb]
\centering
\subfigure[queriy frequency vs. number of terms]{
\includegraphics[width=0.31\textwidth]{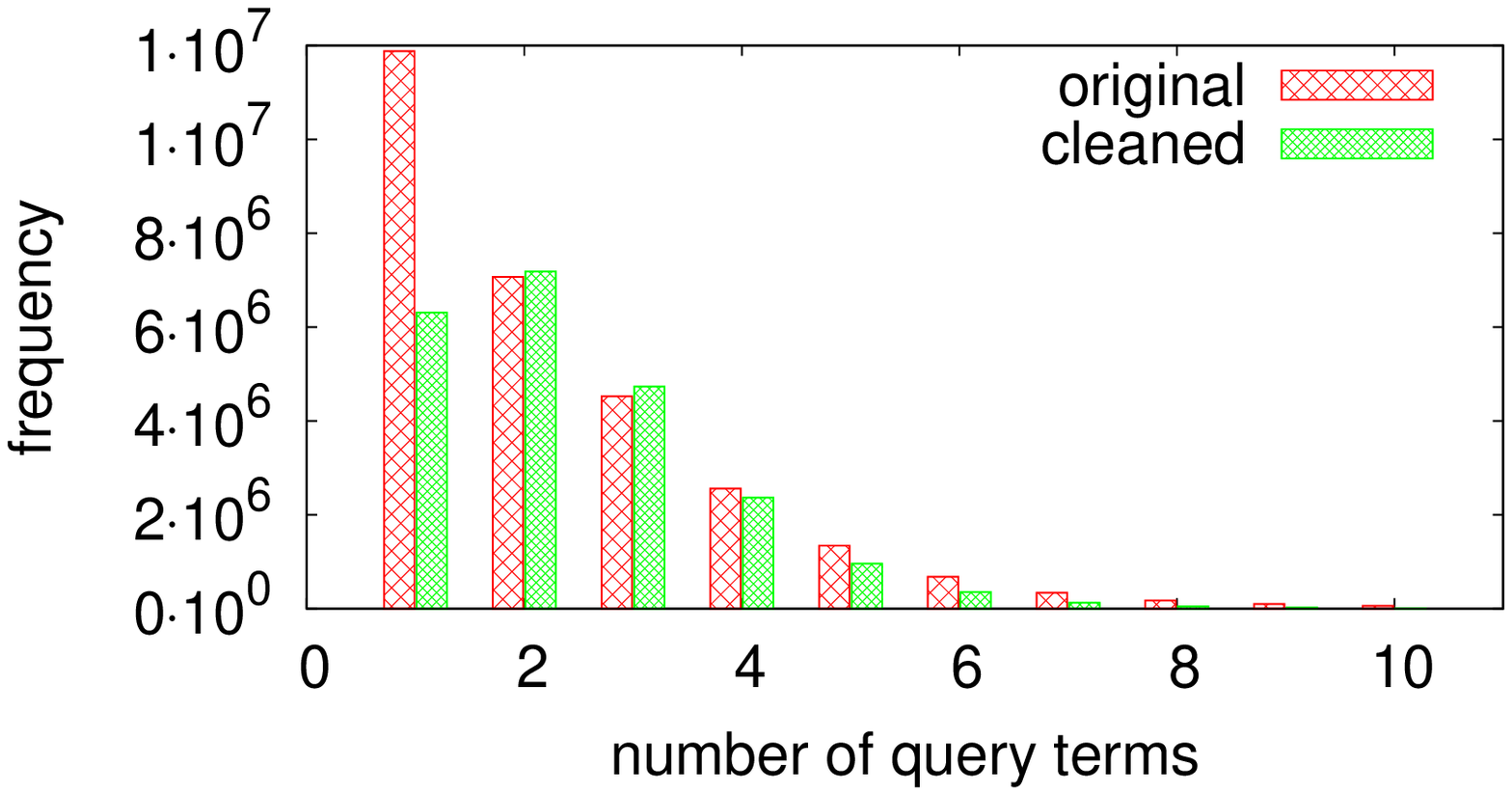}
\label{fig:aol_term_counts}
}
\subfigure[key distribution]{
\includegraphics[width=0.31\textwidth]{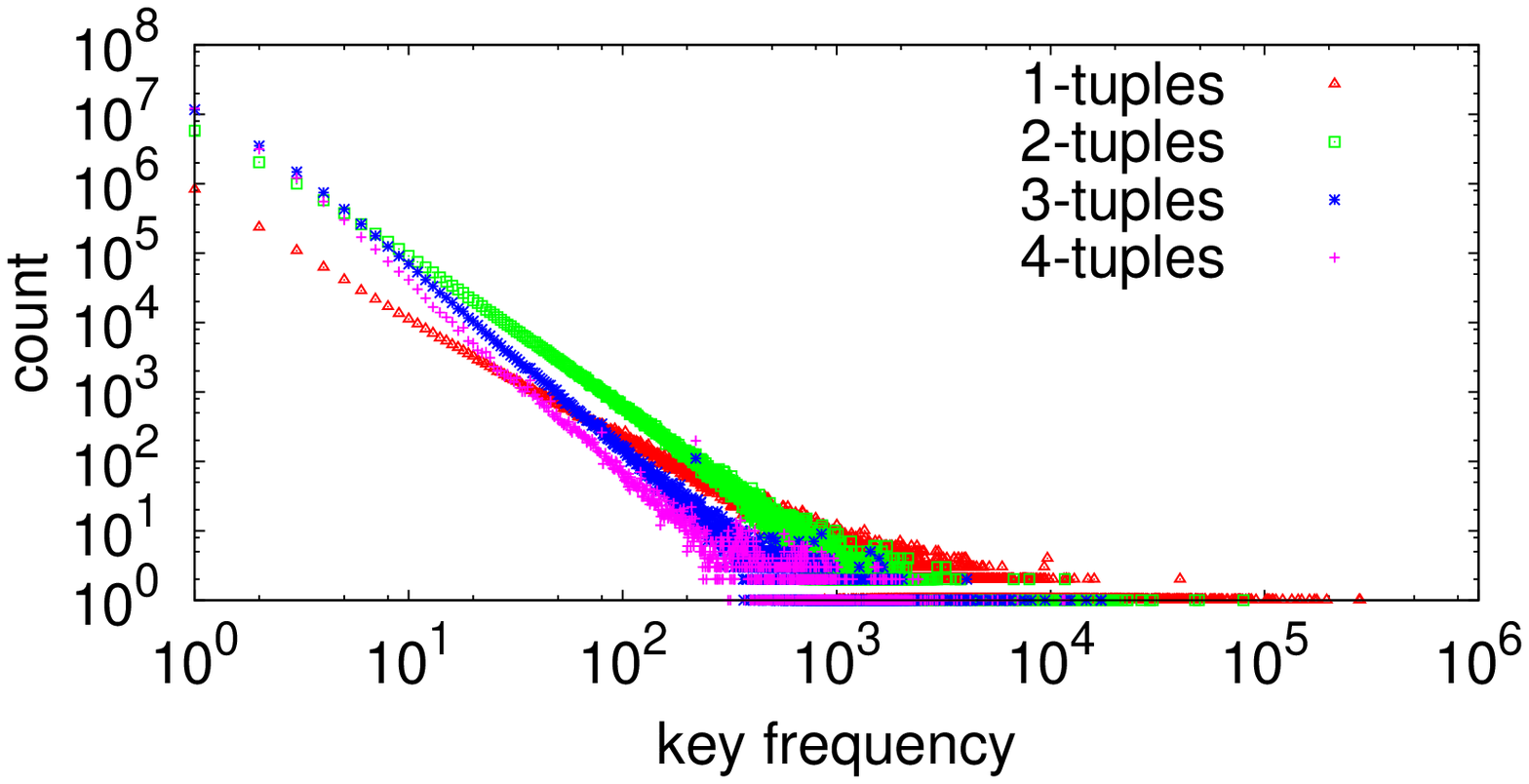}
\label{fig:aol_key_distribution_all}
}
\subfigure[Number of keys for various key sizes]{
\includegraphics[width=0.31\textwidth]{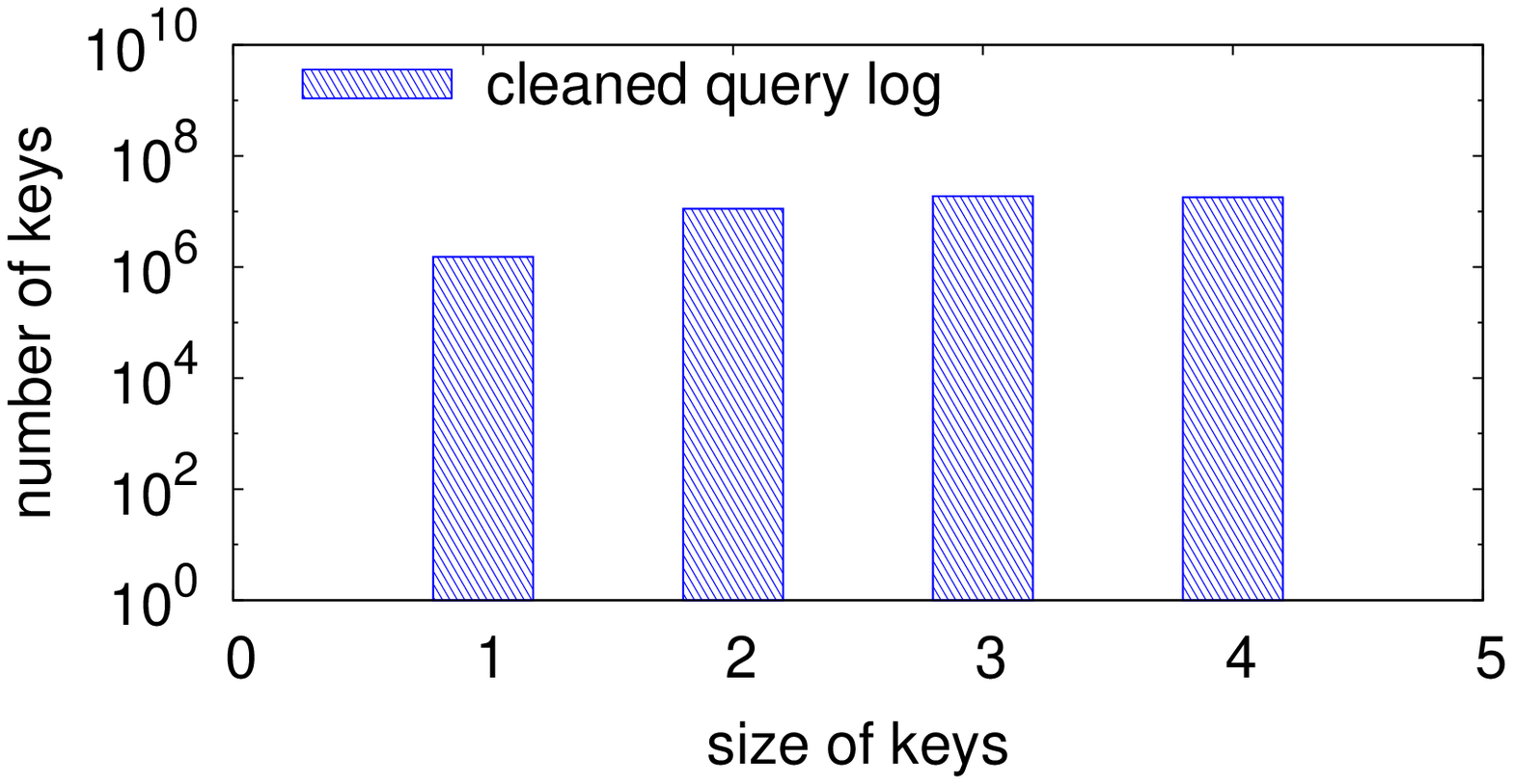}
\label{fig:aol_term_set_distinct_sets}
}
\label{fig:query_log_analysis}
\caption[]{Analysis of cleaned AOL query log.}
\end{figure*}

Summing up, the largest effect on the query log cleaning had the removal of urls, which involved about 1/4 of all queries. However we deem this step reasonable, since the misuse of a search engine as browser address bar is a common phenomenon confined to search engines and we do not expect such behaviour from users when querying a tagging site.
\\
\\
\textbf{Query log analysis.}
The original dataset comprises as set of approximately 28.6 million queries with an average of 2.34 terms per query. After our data cleaning steps the number of queries is approximately 21.0 million, mainly due to the removal of URL query strings. Given the logging period of three months, users have issued 160.3 search requests on average. The average number of terms per query rose slightly to 2.43. Figure~\ref{fig:aol_term_counts} shows the distribution of queries regarding their number terms for 1-10 terms, which reflects 99.6\% of all queries of the original query log and 99.9\% of all queries of the cleaned query log. The most significant difference is the drop of queries with only a single term (removal of URL queries). For two and three terms the number of queries has actually risen after the data cleaning. The reason for this is removal of inappropriate terms (only non-alphanumeric characters, more than 30 characters) from queries with four terms or more.
After the cleaning process more than 73.5\% of all queries comprised more than one term. This result motivates the storage keys of a size large than one in the inverted index in order to avoid the potentially data-intensive computation of the intersections of the single-key inverted lists. Further, only 18.2\%/7.1\% of all queries comprise more than three/four search terms. Thus, setting $s_{max} = 3$ seems to constitute reasonable upper bound for maximum size of keys stored in the inverted index.

Next, like for the tags in the \textsc{Delicious} and \textsc{Flickr} datasets, we looked at the distribution of keys. Here, key refers to the set of query terms of various size, up to $s_{max}$, that can be derived from a query. Analogously to previous tests we computed the keys $k$ of sizes $1 \leq |k| \leq 4$, i.e. for each query we derived all possible keys ($=$ non-empty subsets of query terms) up to size 4. For each key size we plotted the relationship between the frequency a key occurred and the number of the key with the corresponding frequency; see Figure~\ref{fig:aol_key_distribution_all}. Again, a power law relationship clearly dominates, i.e. most keys are unique or rather rare but some keys are quite frequently queried. This is true for all key sizes. Table~\ref{tab:PowerLawQueryLog} shows for each key size the parameters to make the data fit into the power law function $\alpha\cdot f^\beta$ where $f$ is the frequency a key occurred in all queries.
\begin{table}
	\centering
	\small
		\begin{tabular}{|l|c||c|c|c|c|}
			\hline
					\multicolumn{2}{|c||}{\multirow{2}{*}{$\alpha \cdot f^\beta$}} 				& \multicolumn{4}{c|}{size of key} \\
			\cline{3-6}
					\multicolumn{2}{|c||}{}								& 1						& 2 				& 3 				 & 4   \\
			\hline\hline
			\multirow{2}{*}{\textsc{AOL}}		&	$\alpha$	& $8.3\cdot 10^5$			& $5.8\cdot 10^6$ 		&  $1.2\cdot 10^7$	 & $1.2\cdot 10^7$	\\
			\cline{2-6}
						&	$\beta$ & $1.8$			& $1.6$ 		& $1.9$	&		 $2.1$		\\
			\hline
		\end{tabular}
	\caption{Power law parameters for the AOL query log}
	\label{tab:PowerLawQueryLog}
\end{table}
As expected, the number of keys that are frequently queried drop with increasing key size. Comparing the results for the tagging platform datasets (cf. Figure~\ref{fig:both_delicious_all}) we note that the number of unique and rare keys do not significantly increase for larger keys. The reason for this is that the average number of terms per query ($\sim$2.4) is smaller than the average number of tags per resource ($\sim$4). To illustrate this more clearly, Figure~\ref{fig:aol_term_set_distinct_sets} shows the number of distinct keys for each considered key size (fully filled bars).

The previous results show that most keys are uniquely or very rarely queried. Since we store only the inverted list for popular keys that means that we can expect a significant reduction of inverted index size compared to the maximum, i.e. storing of all keys possible depending on the maximum number of considered tags $t_{max}$ and the maximum key size $s_{max}$.

\section{A Multi-Term Based Query Processor}
\label{sec:queryProcessing}
The query retrieval process exploits the current state of the inverted index and cache to answer a query. Particularly with the support of multi-term keys there are several ways to process the query.
In general, as soon as a query comprises two terms or more
(a) not all keys that can be derived from a multi-term query are available in index, and therefore potentially available in the cache and
(b) not all available keys are required to cover all query terms.
\begin{myExample}
Let $q=$\{$t_1$, $t_2$, $t_3$, $t_4$\} be a query containing four terms. With $s_{max}=3$ the following set keys can be derived from $q$; gray marked keys are available in the index:
\begin{center}
	\includegraphics[width=0.48\textwidth]{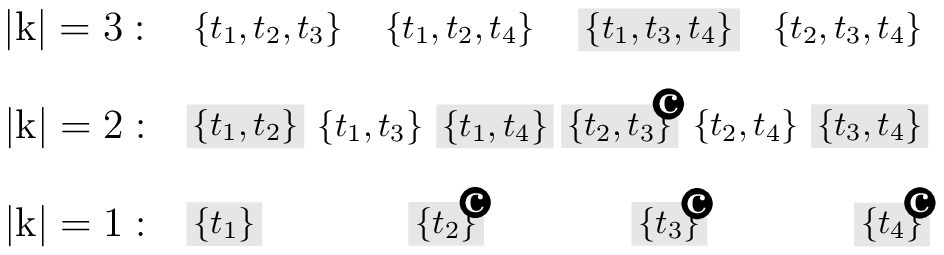}
\end{center}
Possible subsets of available keys to answer query $q$ are, e.g., $\{\{t_1, t_2\}, \{t_1, t_3, t_4\}\}$ or $\{\{t_1\}, \{t_2, t_3\}, \{t_4\}\}$. In contrast, $\{\{t_1, t_2\}, \{t_1, t_4\}\}$ is insufficient since $t_3$ is not covered.
\end{myExample}\\
Given several alternatives to answer a query, identifying the order of keys to eventually access the index and cache with respect to the resulting performance overhead is the most crucial task. The overall goal is to minimize the required bandwidth and the load of nodes. In a nutshell, the most relevant parameters are the length of a key's inverted list and whether a key is cached or not.
In the following we present the involved algorithms for the retrieval process in detail.
\\
\\
\textbf{Retrieval process.} 
If a user issues a query $q$ to any arbitrary gateway node, this node forwards $q$ to gateway node $GN_q$ that is responsible for $q$, i.e. the node responsible for the key derived from $q$. 

\textit{Initiation and basic retrieval process.} On $GN_q$ we then initiate the retrieval process, see Algorithm~\ref{alg:mine}. If query $q$ contains $\leq$ $s_{max}$ terms the key $k_q$ derived from $q$ is potentially available in the inverted index or even in the cache of $GN_q$. Thus, we first access the local cache of $GN_q$ and, if unsuccessful, the back end node that responsible for $q$ (Line~1-6). 
If both attempts to answer $q$ directly fail or the number of query terms is larger than $s_{max}$ we compute all relevant keys (derived from all possible term combinations up to size $s_{max}$) for $q$ (Line~7). 
Next, we retrieve for each available key the size of its inverted list size, again first by trying the access the local cache of $GN_q$, and if that fails by accessing the index (Line~9-12). Note that for each key $k\in K_q^{avail}$ we now, whether $k$ was locally found in the cache or not.
Only if no inverted list has a size of 0, we proceed; otherwise we return an empty result (Line~13-14). 
From the set of available keys $K_q^{avail}$ we then derive the ordered list of keys that eventually specifies the order in which we access the index and cache (Line~15; described in next paragraph). 
Finally, we access the index or cache by sending a process query request to the gateway or back end node responsible for first key in the list (Line~16-17).
\begin{algorithm}[!t]
\footnotesize
\SetKwComment{tcc}{// }{}

\KwIn{query $q = \{t_1, t_2,...,t_n\}$ }

\BlankLine

\If{$|q| \leq s_{max}$}{
		$result \leftarrow $ getFromCache($q$.hashCode()) \;
		\If{$result = null$}{
			$result \leftarrow $ getFromIndex($q$.hashCode());
		}
		\If{$result \neq null$}{
			\Return $result$ \;
		}
}
\BlankLine
$K_q \leftarrow$ computeSubsetKeys($q$) \;
\BlankLine
$K_q^{avail} \leftarrow \emptyset$ \;
\ForEach{$k\in K_q$}{
	$sizes[k]$ $\leftarrow$ getResultSize($k$) \;
	\If{$sizes[k] \neq null$}{
		$K_q^{avail}=K_q^{avail} \cup k $
	}
}\BlankLine
\If{$\exists k \in K_q^{avail}: size[k] = 0$ }{
		\Return $\emptyset$ \;
}
\BlankLine
$L_q^{access} \leftarrow \mathrm{computeKeyAccessList}(q, K_q^{avail})$;\
\BlankLine
\BlankLine
$target = L_q^{access}$[0].hashCode();\quad  $result = \emptyset$ \;
send($target$, $result$, $L_q^{access}$, $q$)\;
\caption{handleQueryRequest(q)}
\label{alg:mine}
\end{algorithm}

If a node receives a process query request we execute Algorithm~\ref{alg:handleList}. 
If the list of keys $L$ is empty -- note that the retrieval ensures that this is is only the case on the gateway node $GN_q$. -- the retrieval process is finished and we can return the result (Line2~1-2); otherwise we proceed. 
To process the current key $k$ we firstly read $k$'s the inverted list, either from the cache in case of a gateway node or from the inverted index in case of a back end node (Line~3).
We then update the intermediate result for query $q$ by computing the intersection between the received intermediate result and $k$'s inverted list (Line~4).
If the intersection and therefore the new intermediate result is empty we can prematurely stop the retrieval process since the final result for $q$ will also be empty. In this case and we remove all keys from $L$; otherwise we only remove $k$ from $L$ (Line~5-8).
After processing $k$, if $L$ is empty the retrieval process is done and we can return the result back to gateway node $GN_q$ using the key derived $q$ as target for the next process query request; if $L$ is not empty, the next target node derives from the new first key in $L$ (Line2~9-12).
As last step, we send the process query request to the new target node (Line~13).

\begin{algorithm}[t]
\footnotesize
\SetKwComment{tcc}{// }{}

\KwIn{target key k, current result, key list $L$, query $q$ }

\BlankLine
\If{$L =  \emptyset$}{
		\Return $result$ \;
}

\BlankLine

$resources \leftarrow \mathrm{getInvertedList}(k)$ \;
$result \leftarrow result \cap resources$ \;
\BlankLine
\If{$result = \emptyset$}{
		$L.\mathrm{removeAll()}$ \;
}
\Else {
	  $L.\mathrm{remove}(k)$ \;
}

\BlankLine
\If{$L =  \emptyset$}{
		$target = q$.hashCode() \;
}
\Else {
	  $target = L_q^{access}$[0].hashCode() \;
}
\BlankLine
send($target$, $result$, $L$, $q$) \;

\caption{handleKeyList(k, result, $L$, $q$)}
\label{alg:handleList}
\end{algorithm}

\textit{Cache and index access strategy.} As mentioned before, there are, in general, various ways to answer a query based on the available keys. Regarding the performance of the retrieval process the goal is to minimize the number of resources, i.e. the entries of the inverted list of keys that have to be transferred within the back end and have to be handled by both the gateway and back end nodes. To find the optimal subset of available keys and their ordering for accessing the cache or index would require complete knowledge, particularly about the expected size of the intersection of two or more partial results. This in turn would require the costly maintenance of comprehensive statistics over the data in the index, which are typically not available, particularly in distributed systems. We therefore propose and discuss in the following a heuristic to determine the set and order of available keys to access the cache and the inverted index.

Algorithm~\ref{alg:FindBestList} implements the heuristic. 
Firstly, we remove all redundant keys from $K_q^{avail}$ (Line~1); a key $k_i \in K_q^{avail}$ is redundant if there is a key $k_j \in K_q^{avail}$ and $k_i \subset k_j$. For example, if $K_q^{avail} = \{k_1=\{t_1, t_4\}, k_2=\{t_1, t_3, t_4\}\}$ we can remove $k_1$ since $k_2$ already covers all terms of $k_1$.
We then generate the list $L$ of keys that specifies the set and order of available keys to access the the cache or inverted index. To minimize the number transferred and handled resources we aim for small intermediate results, and therefore initialize $L$ with the key having the shortest inverted list (Line~3). 
We then iteratively add keys to $L$ that maximize $L$'s coverage of $q$ until $L$ covers $q$, i.e. all terms in $q$ are represented in at least one key in $L$. The rationale for this approach is two-fold. Firstly, maximizing the coverage minimizes the number of required keys to answer q and therefore the number of transfers between nodes. And secondly, keys of larger size tend to have significantly shorter inverted lists than, e.g., single-term keys. If several keys maximize the coverage, we add the one with the smallest partial result. If still no unique key could be identified we favour a random cached key over a random key to add to $L$ (Line~4-10). 
\begin{algorithm}[t]
\footnotesize
\SetKwComment{tcc}{// }{}

\KwIn{set of keys $K$ }

\KwOut{list $L_q$ of keys}
\BlankLine
$K \leftarrow K \setminus \{k_i \in K|\ \exists k_j \in K: k_i \subset k_j  \}$ \;
\BlankLine
$L \leftarrow \emptyset$;\quad  $l \leftarrow L.\mathrm{size()}$ \;
$L.\mathrm{add}(\ \argmin_{k \in K}{size[k]}\ )$ \;
\BlankLine
\While{$\bigcup_{k_i \in L}k_i\neq K$}{
		 $K^{max\_coverage} \leftarrow \argmax_{k \in K}{|k \cup L|}$\;
		 \BlankLine
		 $K^{smallest} \leftarrow \argmin_{k \in K^{max\_coverage}}{size[k]}$ \;
		 \BlankLine
		 $k^{next} \leftarrow K^{smallest}$.getCachedKey() \;
		 \BlankLine
		 \If{k$^{next}$ = null}{
				$k^{next} \leftarrow K^{smallest}$.getRandomKey() \;
			}
		 $L.add(k^{next})$ \;
}
\BlankLine
\For{$i=1$ \textbf{\normalfont \textbf{to}} $l-1$} {
		\If{$\neg L[i\!-\!1]$\normalfont.isCached() \normalfont\textbf{and} $\neg L[i\!+\!1]$\normalfont.isCached()}{
				$L[i]$.unsetCached() \;
		}

}
\BlankLine

\Return $L$ \;

\caption{computeKeyAccessList(q, K)}
\label{alg:FindBestList}
\end{algorithm}

So far, we added keys to $L$ with little concern whether the keys are cached or not; we address this issue in the subsequent discussion. Still, $L$ potentially contains several keys that are cached on the gateway node $GN_q$. In the last step of Algorithm~\ref{alg:FindBestList} we actually exchange the cached copy of a key with the one in the index stored on a back end node (Lines~11-13). To motivate this step, consider the following to key list $L_1=\{...,k_{i-1}, k^\mathbf{c}_i, k_{i+1},...\}$ and $L_2=\{...,k_{i-1}, k_i, k_{i+1},...\}$, where superscript \textbf{c} of a key $k$ indicates that a cached copy of $k$ exists. For both lists, processing the keys $k_{i-1}$, $k_i$ and $k_{i+1}$ results in two transfers between nodes with the same number of resources transferred. In this case we favor $L_2$ to avoid additional load for the gateway nodes. In other words, we only send a process query request to $GN_q$ if the retrieval process is done or it indeed reduces the number of transferred resources. However, this step is only relevant if $|L|>5$, thus for queries with at least five query terms.
\\
\\
\textbf{Cost analysis and discussion.}
Regarding the required bandwidth, the share that is specific to our multi-term approach is the consideration of all relevant keys up to size $s_{max}$ for a given query $q$ to access the inverted index; see Algorithm~\ref{alg:mine}, Lines~7-12. In the worst case, no relevant key is locally cached on the gateway node handling $q$. In this case the algorithm performs $\left [ \binom{|q|}{1} + \binom{|q|}{2} + ... +  \binom{|q|}{s_{max}} \right ] \in O(|q|^{s_{max}})$ accesses to the index. In practice, however, this polynomial growth has only a limited impact on the performance. Firstly, as our analysis shows, the value for $|q|$ is rather small ($\sim$2.4 on average) and a reasonable value for $s_{max}$ is with 3 or 4 also small. Secondly, the $O(|q|^{s_{max}})$ index accesses are only required to retrieve the length of the corresponding inverted lists, and not the lists themselves. The actual size of the data transferred, e.g. in terms of required bandwidth, is thus very small.

When computing the key list to access the cache and index using Algorithm~\ref{alg:FindBestList} we emphasize more on the size of the relevant key's inverted list and less on the fact whether the keys are cached. With that, there are cases conceivable in which Algorithm~\ref{alg:FindBestList} does not return the key list. To give an example, removing all redundant keys might remove the set of keys that would otherwise allow answering a query completely using the cache. The main reasons for our decision are:
\\
(1)~For single-term queries this issue has no effect on the performance. Given a single-term query $q_s$ only the one key $k_{q_s}$ derived from $q_s$ is relevant to answer $q_s$. If $k_{q_s}$ is cached then it is so on the gateway node handling $q_s$. Additionally, single-term queries have a significant impact on the overall performance since they still pose a large number of user queries and feature, in general, a much larger inverted list than multi-term keys.
\\
(2)~Algorithm~\ref{alg:FindBestList} minimizes the number of keys accessed to answer a query. Not removing redundant keys, e.g. in order to increase the number of cached keys, results in key list containing more keys and/or at least keys with a larger inverted lists. Further, from our data analysis and our evaluation we observe that cached keys tend to feature a higher-than-average long inverted list. The reason for this is that frequent tag combinations are also likely to represent frequent term combinations in user queries. We therefore, in order to limit the additional load for the gateway nodes, favour the handling and transferring of smaller data in the back end over the handling of larger data on the gateway nodes.

\section{Index and Cache Maintenance}
\label{sec:Index}
In this section we describe our approach for a query-driven maintenance of the inverted index on the back end nodes and the cache on the gateway nodes in detail.

\subsection{Maintenance of Inverted Index}
The maintenance of the inverted index comprises two major tasks: suspending and resuming of keys depending on their popularity and the handling of updates on the tag data.
\\
\\
\textbf{Suspending and resuming keys.}
\label{subsec:indexSuspendingResuming}
The inverted index stores only the inverted lists of popular keys, where the popularity of a key $k$ is derived by the frequency how often $k$ is requested during query processing. If a key $k$ becomes unpopular, we \textit{suspend} $k$, i.e. we delete $k$'s inverted list and mark $k$ as unavailable for processing queries. As soon as a suspended or new key $k$ becomes popular, we \textit{resume} $k$. Resuming a key $k$ involves retrieving its corresponding inverted list which in turn translates to performing a query for $k$ (cf. Algorithm~\ref{alg:mine}) and storing the result as $k$'s inverted list. As last step, we mark $k$ as available again.

To measure the popularity, we provide each key $k$ with a bit vector $\mathcal{B}_k$ of length $\ell$. Every time $k$ is requested, we first set $\mathcal{B}_k := \mathcal{B}_k >> 1$, i.e. we shift the bit vector for $k$ one bit to the right, and then set $\mathcal{B}_k := \mathcal{B}_k\ |\ 2^\ell$, where operator $|$ performs a bitwise inclusive $OR$ operation.
Further, to implement the timely decay of a key $k$'s popularity, we periodically, after time $\Delta^{decay}$, set $\mathcal{B}_k := \mathcal{B}_k >> 1$. With that, the number of set bits in $\mathcal{B}_k$ represents the popularity of a key $k$.
\begin{myExample} The following figure shows a bit vector $\mathcal{B}_k$ both after a request for $k$ and after a periodically shifting.
 \begin{center}
	\includegraphics[width=0.40\textwidth]{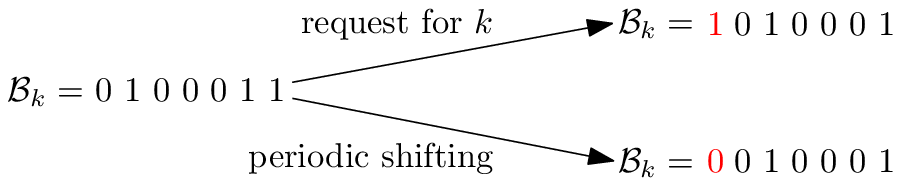}
\end{center}
While each periodic shifting decreases the number of set bits, a request for $k$ increases the number or keeps it.
\end{myExample}

Beside vector length $\ell$ and interval $\Delta^{decay}$, further relevant parameters are 
(a) $b^{res}$ as the minimum number of set bits in $\mathcal{B}_k$ to resume $k$ and 
(b) $b^{susp}$ as the number of set bits in $\mathcal{B}_k$, when falling below, to suspend $k$. 
To be meaningful, i.e. so that non-empty set of popular multi-term keys are actually indexed, $b^{susp} < b^{res}$ must hold.
Resuming keys adds to the workload for processing user queries. However, depending on the choice of the values for these four parameters, we expect resuming keys to be much more infrequent events than evaluating user queries.
\\
\\
\textbf{Handling updates of tags.}
Updating a resource (here, a web page) by adding or deleting a tag must be propagated to the inverted index. A na\"{i}ve way to do so is that the node responsible for storing an updated resource sends an update message to each relevant key which the update affects. In case of a newly added or deleted tag $t$, the relevant keys comprise all keys that can be derived from the set of available tags before adding $t$ or after deleting $t$ in combination with $t$ itself, up to size $s_{max}$. 
The number of relevant keys, and therefore the number required update messages for a single update depends on the current set of indexed tags of a page $p$. Since the number if $p$'s indexed tags $\leq t_{max}$, the number of messages is in $O({t_{max}}^{s_{max}})$. With this approach the index is always up to date. 
However, although the size of an update message is small, the number of messages per update is very large. Thus, while this approach performs well in systems with infrequent updates, it is not suitable for high update rates like we observed in \textsc{Delicious} and \textsc{Flickr}. Further, the node storing a resource is not aware of suspended keys in the index, but sends messages to all relevant keys. Thus, a potentially large number an unnecessary update messages are sent to suspended keys.

To guarantee that the inverted index is always up to date inherently requires the immediate and costly propagation of each update to all relevant keys. We therefore propose an update mechanism which relaxes the guarantee of the timeliness of the index but resulting in a significant decrease of bandwidth consumption. In a nutshell, we propagate the information about a new or deleted tag only to the corresponding single-tag key in the inverted index. We further update only available multi-term keys periodically. To do so, we propose \textit{incremental update queries}, where the results only contain the relevant changes, i.e. the tags to be added or to be deleted, for a multi-term key's inverted list. In the following, we present our update mechanism in detail.

\textit{Extensions to the inverted index.} When a user adds or deletes add tag $t$ from a resource $r$, the node storing $r$, sends an update message to the single-term key $k$ representing tag $t$. The update message contains $r$ and the information whether $r$ is to be added or to be deleted. In case of a new resource, we add $r$ to $k$'s inverted list; however, we do not immediately remove resources from a key's inverted list but only mark them as deleted.
If a new resource is already in the list but marked as deleted, we simply unmark the resource.
To support incremental update queries to update multi-term keys, nodes have to distinguish between resources that have already been propagated to multi-term keys and both newly added and deleted resources. To accomplish this, we assign a timestamp to each resource in the inverted list of single-term keys, indicating when it has either been added or marked as deleted. Secondly, we assign a timestamp to each multi-term key, indicating the time of its last update. Thus, for a multi-term key $k_m$, we can identify all resources in the inverted lists of all single-term keys $k_i$, $\forall i: k_i \subset k_m$, that have been added or deleted after the last update of $k_m$.
\begin{myExample}
In the figure below, superscripts represent the timestamps for resources (time when added or marked as deleted) and keys (last update). Crossed out resources are marked as deleted.
\begin{center}
		\includegraphics[width=0.30\textwidth]{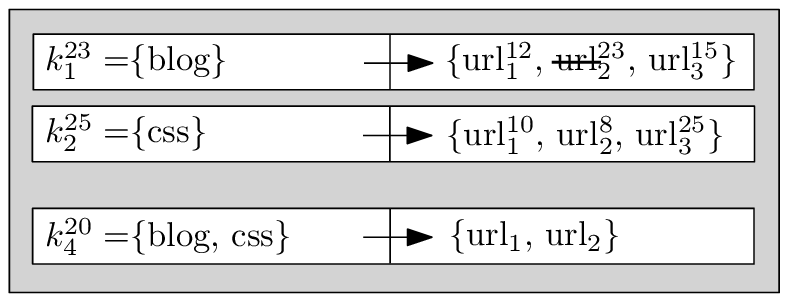}	
\end{center}
Key $k_4$ represents the intersection of $k_1$ and $k_2$ at time $t = 20$. After that time, a user has deleted the tag ``blog'' from $url_1$ (at $t=23$) and added the tag ``css'' to $url_3$ (at $t = 25$).
\label{exp:extendedIndex}
\end{myExample}

That resources are marked as deleted but not immediately removed from inverted lists is due to our goal to avoid using the complete inverted lists of single-term keys to update multi-term keys (as described in the next paragraph). To guarantee that all updates of multi-term keys are correct, we have to ensure that no marked resource is removed from an inverted list of a single-term key $k_s$, before all available keys $k_i$ that contain $k_s$, i.e. $\forall i: k_s \subset k_i$, have been updated. To accomplish this, we define $\Delta^{update}$ as the maximum period of time before updating a multi-term key. Thus, after a time of $\Delta^{update}$, starting from the time a resource $r$ has been marked as deleted, we can safely delete $r$ from the inverted list.

\textit{Incremental updates of keys.} Basically, we can update a multi-term key $k_m$ by simply issuing a query for $k_m$ using Algorithm~\ref{alg:mine}. (Note that we would have to make a minor modification so that the algorithm only requests single-tag keys to process the query.) However, this approach would result in an unnecessary bandwidth consumptions, since, in general, the result of such a na\"{i}ve update query would contain mostly resources that are already covered by $k_m$. Incremental update queries exploit this fact. The basic idea is to only transfer the latest changes in the inverted lists of single-term keys to evaluate the necessary changes required to update multi-term keys. Latest changes in an inverted list refer to the set of resources added or marked as deleted after the last update of a multi-term key.

\begin{figure*}
	\centering
		\includegraphics[width=.80\textwidth]{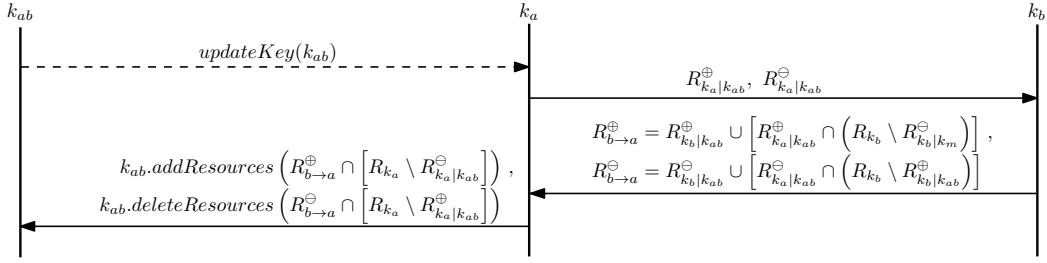}
	\caption{Incremental update for a two-term key $k_{ab}$. The inverted list of $k_ab$ is the intersection of the inverted lists of both single-term keys $k_a, k_b \subset k_{ab}$}
	\label{fig:incremental_update}
\end{figure*}

\begin{figure}[thb]
	\centering
		\includegraphics[width=0.49\textwidth]{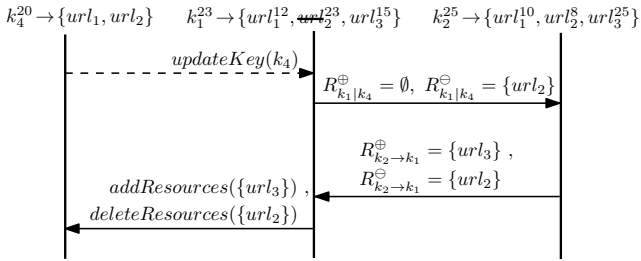}
	\caption{Example of an incremental update of a two-term key}
	\label{fig:incremental_update_example}
\end{figure}

To more formalize the concept of incremental update queries, let $R_{k_i}^{\ominus}$ be the set of resources in the inverted list of key $k_i$ that are marked as deleted; $R_{k_i}^{\oplus}$ contains all resources not marked. Further, let $ts(r)$ be the timestamp when a resource was added or marked as deleted in an inverted list, and $ts(k)$ the timestamp of the last update of a key $k$. We then can define $R_{k_i|k_j}^{\oplus} = \{ r \in R_{k_i}^{\oplus}|\ ts(r) > ts(k_j) \}$ as set of added resources in $k_i$'s inverted list with a timestamp older than the timestamp of a key $k_j$; analogously we define $R_{k_i|k_j}^{\ominus} = \{ r \in R_{k_i}^{\ominus}|\ ts(r) > ts(k_j)\}$. With these definitions, Figure~\ref{fig:incremental_update} shows the involved steps for an incremental update of a two-term key. Additionally, Example~\ref{exp:incrementalUpdate} shows the update process for a small index data set.
\begin{myExample}
Figure~\ref{fig:incremental_update_example} illustrates the update of key $k_4$ from Example~\ref{exp:extendedIndex} at a time $t\!>\!25$, e.g. $t\!=\!30$. The squence diagram shows the sets of resources that are sent between the nodes storing the involved keys. After adding $url_3$ and deleting $url_2$ the inverted list of key $k_4$ is $k_4^{30}\rightarrow \{url_1, url_3\}$.
\label{exp:incrementalUpdate}
\end{myExample}

\begin{figure}[thb]
	\centering
		\includegraphics[width=0.35\textwidth]{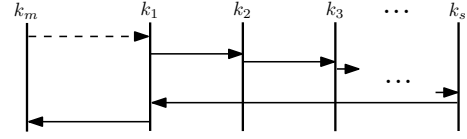}
	\caption{Schematically illustration of the incremental update of a multi-tag key $k_m$ with size $s$}
	\label{fig:incremental_update_extended}
\end{figure}

Extending the update process for keys of size $> 2$ is straightforward. Consider a multi-tag key $k_m$ of size $s$ and the corresponding single-tag keys $k_1, k_2, ..., k_s \subset k_m$. The basic mechanism is that the changes of each $k_i$'s inverted list are successively incorporated into the intermediate results, before eventually sent back to update $k_m$.
Figure~\ref{fig:incremental_update_extended} schematically illustrates the transfer of the intermediate update results along the chain of single-tag keys $k_i \subset k_m$. Note that there is no pre-defined order in which the single-tag keys update the intermediate results.
\\
\\
\textbf{Cost Analysis}
The query-driven maintenance and particularly the support of updates adds further load, both in terms of storage and processing power, to the basic one for evaluating user queries. We now analyze first the required additional storage, followed by a performance analysis regarding the concept of incremental update queries.

\textit{Storage requirements.}
The additional storage requirements for the support of incremental update queries are twofold. Firstly, we cannot immediately remove deleted resources from the inverted list of single-tag keys. Secondly, all keys and all the resources in inverted lists of single-tag keys require a timestamp. Since newly added resources are not specific to the incremental update process, only the number of resources that are marked as deleted in the inverted lists of single-tag keys add to the required storage. This number depends on the frequency $f_{act}$ of user actions (adding or deleting tags) and the maximum period of time $\Delta_{max}$ before updating a multi-tag key. Further, we can distinguish between the average frequency of adding tags $f_{act}^\oplus$ and deleting tags $f_{act}^\ominus$, where $f_{act} = f_{act}^\oplus + f_{act}^\ominus$. With that the number of resources marked as deleted in inverted lists of single-tag keys is in $O(\frac{f_{act}^\ominus}{f_{act}^\oplus + f_{act}^\ominus}\cdot \Delta_{max})$. The datasets we analyzed only allows to give the numbers for $f_{act}^\oplus$: $34.32\frac{actions}{minute}$ for \textsc{Delicious} and $107.11\frac{actions}{minute}$ for \textsc{Flickr}. To give some absolute numbers, we assume the worst case, i.e. (a) all user actions are deleting tags and (b) all single-tag keys are available. Table~\ref{tab:StorageIncrease} shows the estimated number of resources marked as deleted for both tagging datasets. The inverted index also comprises the inverted lists of available multi-tag keys. Thus, the overall ratio of marked resources is smaller than the values given in Table~\ref{tab:StorageIncrease}. However, the number of available multi-tag keys is hard to quantify, due to their their query-driven suspending and resuming. In addition, in real-world systems we expect much lower values, since we assume that adding tags is much more frequent than deleting tags, reducing both the absolute and particularly the ratio of resources marked as deleted.
\begin{table}[t]
	\centering
		\begin{tabular}{|l|r|r|r|r|}
			\hline
																& 1 hour					& 1 day 				& 1 week 				& 1 month   \\
			\hline\hline
			\textsc{Delicious}				& 0.02\%			& 0.39\% 		& 2.7\%		& 12.0\%		\\
			\hline 
			\textsc{Flickr} 					& 0.03\%			& 0.69\%		&	4.9\%		&	21.5\%		\\
			\hline
		\end{tabular}
	\caption{Estimated ratio of resources marked as deleted in inverted lists of single-tag keys for various $\Delta_{max}$}
	\label{tab:StorageIncrease}
\end{table}


\textit{Bandwidth consumption.}
We evaluate the required bandwidth costs in the worst case. For the following analysis we consider the update of an arbitrary multi-tag $k_m$ of size $s$. Thus, an incremental update query refers to $s$ single-tag keys $k_1, k_2, ..., k_s \subset k_m$. Regarding the required bandwidth, the worst case occurs when each tag, newly added or marked as deleted in any inverted list of a single-tag key $k_i \subset k_m$, results in a distinct update of the current inverted list of $k_m$. 
To be more precise, each newly added resource $r^\oplus$ in the inverted list of key a $k_i \subset k_m$ must already be present in inverted lists of all other keys $k_j$ and not marked as deleted, i.e. $\forall k_j \subset k_m, k_j\neq k_i: r^\oplus \in (R_{k_j} \setminus R_{k_j|k_m}^\ominus)$. Analogous, each marked as deleted resource $r^\ominus$ in the inverted list of key $k_i$ must already be present in inverted lists of all other keys $k_j$ but must not be an newly added tag in each other inverted lists, i.e. $\forall k_j \subset k_m, k_j\neq k_i: r^\ominus \in (R_{k_j} \setminus R_{k_j|k_m}^\oplus)$. In all other cases, a single update will get ``lost'' in the incremental update process, which in turn reduces the overall size of transferred data.
	
In this worst case scenario node $n_{k_1}$ storing $k_1$ sends $\sum_{i=1}^1(|R_{k_i|k_m}^\oplus|+|R_{k_i|k_m}^\ominus|)$ resources to the node $n_{k_2}$ storing $k_2$, $p_2$ sends $\sum_{i=1}^2(|R_{k_i|k_m}^\oplus|+|R_{k_i|k_m}^\ominus|)$ resources to node $n_{k_3}$ storing $k_3$ and so forth. Thus, the number of transferred resources by node $n_i$ is 
\begin{equation} 
  TR_{n_i} = \sum_{j=1}^{i}(|R_{k_j|k_m}^\oplus|+|R_{k_j|k_m}^\ominus|)
\end{equation}	
With that the total number of transferred resources $TR_{total}$ sent during an incremental update is
\begin{eqnarray} 
  TR_{total} & = & \sum_{i=1}^{s} { \sum_{j=1}^{i}(|R_{k_j|k_m}^\oplus|+|R_{k_j|k_m}^\ominus|)} \\ \nonumber
  & &\ +\ \sum_{j=1}^{s}(|R_{k_j|k_m}^\oplus|+|R_{k_j|k_m}^\ominus|)
\end{eqnarray}	
The latter summand represents the number of resources node $n_{k_1}1$ eventually sends to node $n_{k_m}$ responsible for $k_m$.
Further, let $r^{max}$ denote the largest number of changes in the inverted list, i.e.
$\forall i, 1\leq i \leq s: r^{max} \geq (|R_{k_i|k_m}^\oplus|+|R_{k_i|k_m}^\ominus|)$. 
Finally, we can specify the upper bound for the number of resource transferred resources sent during an incremental update $TR_{max}$ as
\begin{equation}
TR_{max}\leq r^{max}\left[ s\cdot \frac{s(s+1)}{2}+2s \right] = r^{max}(s^2+3s)
\end{equation}
Two points are worth mentioning. Firstly, since $s \leq s_{max}$, the value for $s$ is rather low, e.g. 3 or 4. Secondly, the worst case scenario for an incremental update requires a lot of conditions to be true and is therefore extremely unlikely. Typically, only a small subset of changes in a single-tag key, if any, yield an update in a related multi-tag key, resulting in a significantly reduction of required bandwidth.

\subsection{Maintenance of Cache}
On an abstract level, the main task for the cache maintenance relate to the ones for maintaining the inverted index, i.e. the insertion and deletion of keys depending on their popularity and dealing with updates on the underlying tag data. However, since the caching layer resides on top of the inverted index, cache maintenance adds only limited complexity. 
Caching keys basically involves storing a copy of available keys on the gateway nodes depending on their popularity. With, in general, several gateway nodes in the system on which node(s) to store the copy of an indexed key, the question of how to distribute the cache arises. Throughout the paper we consider the following two extreme cases:

\textit{Uniform caching.} 
With uniform caching, the cache is replicated among all gateway nodes, i.e. each node stores all cached keys. Regarding the number of cache hits this straightforward approach yields the optimal case. Whichever gateway node handles as query, has instant access to the full cache. However, storing the complete cache on all gateway nodes has a negative impact on the required cache maintenance. Both newly popular keys (incl. their inverted list) and updates must be propagated to all gateway nodes.

\textit{Dedicated caching.}
In this setting, the cache is distributed among all gateway nodes, each cached key stored on one dedicated gateway node. This minimizes the maintenance overhead due to the insertion of keys and the propagation of updates.
To improve the matching between the shares of the cache a gateway nodes stores and the queries it handles we exploit the DHT-like organization of the gateway nodes. That means, we store (a) a cached key on the gateway node that is responsible for $k$ and (b) forward a query $q$ to the gateway node that is responsible for the key derived from $q$. With that, all single-term queries are handled by the ``correct'' gateway node, i.e. the one which potentially stores the corresponding key.

Obviously, various combinations between uniform and dedicated caching are conceivable. For example, one might apply dedicated caching for single-term keys and uniform caching for multi-term keys. Another alternative is to store all cached (multi-term) keys on a selected subset of gateway nodes. However, for the sake of clarification, we strictly distinguish in our evaluation between uniform and dedicated caching.
\\
\\
\textbf{Insertion and deletion of keys.}
Similar to the inverted index, we cache a key $k$ depending on its popularity, again derived from its bit vector $\mathcal{B}_k$. We define $c^{ins}$ as minimum number of set bits in $\mathcal{B}_k$ to cache key $k$. Since we cache only keys that are available in the index, $b^{res} \leq c^{in} \leq \ell$ must hold. Note that $c^{ins}$ addresses both single-term and multi-term key, in contrast to the index where we always store the inverted lists of single-term keys. Analogously, $c^{del}$ denotes the number of set bits to delete a key from the cache. Besides the reasonable condition $c^{del} < c^{ins}$, also $c^{del} \geq b^{susp}$ must hold to ensure that only available keys are in the cache.

If the number of set bits in $\mathcal{B}_k$ of an index key $k$ exceeds $c^{ins}$, the back end node $n_b$ responsible for $k$ caches $k$ according to applied caching scheme. In case of dedicated caching, $n_b$ forwards a copy of $k$ and its inverted list to the gateway node responsible for $k$; $n_b$ sends the copy to all gateway nodes in case of uniform caching. Once a the number of set bits in $\mathcal{B}_k$ are $\leq c^{del}$, the back end node $n_b$ storing $k$ sends an request to the corresponding gateway node(s) -- depending on the caching scheme -- to delete $k$ from the cache. As a consequence of this approach for inserting and deleting keys, a back end node can keep track which of its locally indexed keys is currently cached.
\\
\\
\textbf{Propagation of updates.}
When considering the propagation of updates we again distinguish between single-term and multi-term keys, however with less effect on involved mechanisms particularly for updating multi-term keys. For a single-term key $k_s$, the back end node storing $k_s$ simply forwards each added or deleted resource in $k_s$'s inverted list to the corresponding gateway node(s). Thus, like for the inverted index, the cached inverted lists of single-term lists are always up to date. Further, compared to the transferred and locally handled resources for processing queries, we expect the additional cost to be are very small.

To update the multi-term keys in the cache we exploit our mechanism of incremental updates and the fact that each back end node knows which of its maintained keys are currently cached. After a back end node $n_b$ updates an indexed multi-term key $k_m$ by means of an incremental update, the copy of $k_m$ in the cache is only affected if the incremental update of $k_m$ yielded a non-empty result. If the result of an incremental update of $k_m$ is not empty, $n_b$ forwards only this result to the gateway node(s) caching $k_m$. The gateway node(s) then incorporate the result of the incremental update in the cached inverted list of $k_m$, i.e. removing or adding the resources representing the difference between inverted list of $k_m$ before and after the update. Like for updating single-term keys, we expect only small performance overhead due to the propagation of updates to the cache. The reason is that the results of incremental updates are -- if non-empty -- tend to be small. Our evaluation in Section~\ref{sec:Evaluation} confirms our expectations.

\section{Evaluation}
\label{sec:Evaluation}
We next report the performance of our multi-term inverted index using a key-value store as distributed back end infrastructure for \textsc{GutenTag} based on trace-driven experiments.
\\
\\
\subsection{Prelimiary Steps and Evaluation Method}
\label{subsec:eval_pre}
So far we have analyzed the datasets from both tagging platforms and the query log separately. The results give indications to the expected characteristics of global multi-tag index -- overall size, number of key, length of inverted lists -- from a dataset and query log perspective. For our evaluation in Section~\ref{sec:Evaluation}, where we investigate the effect of our indexing scheme by means of a prototypical retrieval engine, we have to consider the tag data and query log in unison. 

However, the meaningful application of a search engine query log on a tagging system is not trivial. Although, due to lack of data, relevant results are still missing, we argue that there are some fundamental differences between searching on a tagging platform and searching the web by means of a search engine. In general, tagging systems allow users to search resources by clicking on existing tags, e.g. using tag clouds. Further, tagging platforms typically show all (popular) tags for a resource. \textsc{Delicious}, for example, also displays related tags, and \textsc{Flickr} allows users to create groups and assign tags to these groups. Users then can navigate over the groups to find more related or further relevant tags. Such features help users to quickly identify ``good'' (existing and relevant) tags for their queries, additionally leading to a rather limited pool of search terms. In contrast, web search engines, in general, do not provide such kind of guiding mechanisms for users. As a result, the diversity of used query terms can be expected to be much wider. We therefore re-construct the AOL query log in two basic steps. Firstly, we remove the queries and query terms from the query log we generally deem inappropriate for a tagging system. And secondly, we create individual query logs for both the \textsc{Delicious} and \textsc{Flickr} derived from the original \textsc{AOL} query log.
\\
\\
\textbf{Matching the vocabulary.} In a first step, we computed the intersection between all distinct terms in the query log and the distinct terms in each dataset. To keep matters simple and consistent we compared all terms and tags using a straightforward string exact-matching. Thus, we do not consider spelling errors, terms in different language (e.g., ``italia''  vs. ``italy''), abbreviations (e.g., ``newyorkcity'' vs ``nyc''), etc. 
For both datasets the intersection, compared to the union of all query terms and tags, is rather small: 7.24\% for \textsc{Delicious} and 7.57\% for \textsc{Flickr}. We then removed all terms of the intersection from the set of query terms and subsequently removed all queries that no longer comprised any term that was not an available tag. Table~\ref{tab:IntersectResultsQueryLog} shows the overall effects of these removal on the so newly dataset-specific generated query logs.
\begin{table*}
\parbox{.30\linewidth}{
\centering
	\footnotesize
		\begin{tabular}{|l|c|c|}
			\hline
			\multicolumn{1}{|c|}{} & \multicolumn{2}{c|}{query log +}				 \\
			\cline{2-3} 
																				& \textsc{Delicious} 			& \textsc{Flickr} \\
			\hline\hline 
			distinct terms 				 						& 10.93\% 			& 10.64\%		\\
			\hline 
			term set													& 90.6\%					&	87.76\%			\\
			\hline
			query set 												& 91.48\%				& 89.13\%			\\
			\hline
			terms / query								  	& 2.41						& 2.40					\\
			\hline
		\end{tabular}
	\caption{Effects of vocabulary matching on query log.}
	\label{tab:IntersectResultsQueryLog}}
\hfill
\parbox{.30\linewidth}{
\centering
	\footnotesize
		\begin{tabular}{|l|c|c|}
			\hline
			\multicolumn{1}{|c|}{} & \multicolumn{2}{c|}{query log +}				 \\
			\cline{2-3} 
																				& \textsc{Delicious} 			& \textsc{Flickr} \\
			\hline\hline 
			distinct tags 				 						& 17.66\% 			& 20.76\%		\\
			\hline 
			tag set														& 78.34\%				&	78.24\%			\\
			\hline
			query set 												& 88.23\%				& 92.47\%			\\
			\hline
			terms / query 								 	& 3.39					& 3.72				\\
			\hline
		\end{tabular}
	\caption{Effects of vocabulary matching on tag datasets.}
	\label{tab:IntersectResultsDatasets}}
\hfill
\parbox{.30\linewidth}{
\centering
\footnotesize
		\begin{tabular}{|l|c|c|}
			\hline
			\multicolumn{1}{|c|}{} & \multicolumn{2}{c|}{query log +}				 \\
			\cline{2-3} 
																				& \textsc{Delicious} 			& \textsc{Flickr} \\
			\hline\hline 
			distinct terms 				 						& 7.27\% 			& 6.70\%		\\
			\hline 
			term set													& 30.96\%			&	16.28\%			\\
			\hline
			query set 												& 42.37\%			& 27.30\%			\\
			\hline
			terms / query									 	& 1.78				& 1.45				\\
			\hline
		\end{tabular}
		\caption{Effects of dataset-adjusted filtering on query log.}
		\label{tab:AdjustedQueryLog}
}
\end{table*}

The difference between the result for \textsc{Delicious} and \textsc{Flickr} are only marginal. In both cases, less than 11\% of all query terms have counterpart by means of a tag in the datasets. However, when looking at the complete term sets of queries, approximately 90\% of all terms is not affected by the removal. This is also true for about the same number of queries. These results are interesting, since they show that the intersections of all query terms and set of \textsc{Delicious} and \textsc{Flickr} respectively cover most of the query terms in the query log. In other words, approximately 10\% of all query terms cover approximately 90\% of all queries. Further, we computed the average number of terms per query for each new dataset-specific query log. Again, in both cases, this value drops only very slightly below 2.43, being the value for the basic query log. The main reason for this is that we have removed all queries without terms, so that these queries are no longer considered in the computation of the average number of terms per query.

In a similar fashion, we analyzed how the rather small set of intersecting terms/tags affects the tag datasets. The relevant questions here are -- instead of the actual removal of tags or resources without tags -- how popular the tags of the intersection are in the actual dataset or how many resources are no longer be addressed by any query. Table~\ref{tab:IntersectResultsDatasets} shows the result. 
Overall, again the results are quite similar for \textsc{Delicious} and \textsc{Flickr} and are also qualitatively similar to the results for the query log. Here, roughly 20\% of distinct tags cover almost 80\% of the complete tag sets for all resources. Further, approximately 90\% of all resources features at least a tag that is relevant for at least one query in the corresponding dataset-specific query log. The average number of tags per resource drops more significantly than the average number of terms per query, from 4.19 to 3.73 for \textsc{Delicious} and from 4.01 to 3.39 for \textsc{Flickr}. The explanation for this is that a lot of resource feature several tags that are no longer addressed by any query.

Summing up, matching the vocabulary has only a small effect on both the query log and the tag datasets. This is due to the fact, the intersection of query terms and tags comprise the most popular terms in the query log and the most popular tags in \textsc{Delicious} and \textsc{Flickr} dataset. In this sense the query log and tag datasets are more related than we have anticipated beforehand. This result would be even more pronounced if one would apply more sophisticated methods when determining the intersection of query terms and tags, like, e.g., consideration of types, synonyms or alternative spellings of the same concept.
\\
\\
\textbf{Removing non-empty queries.} Matching the vocabulary brought the query log and the tagging datasets much closer together without sacrificing their basic characteristics, e.g., size, average number of terms/tag in query log/dataset. However, we noticed that a rather large number of queries still result in empty results. Thus, many queries feature a combination of terms while no resource features the same corresponding combination of tags (although each term for itself can be found as tag). While this is not an actual problem, it might have a significant impact on our evaluation. With many multi- term queries yielding empty result, we could (a) potentially answer a lot queries with a minimum of transferred data and (b) the global index would to a large portion contain keys with an empty inverted list. Since tagging platforms provide various means (showing related tags or all available tags of a resource) to help user refining their queries, we do not expect a significant number of empty query results. Thus, the current dataset-specific query logs with only a matching vocabulary would be rather unrealistic and could unduly distort the results of our evaluation in our favour. In order to prevent such possibly biased optimistic results, as an alternative approach, we extract all queries that return non-empty results, which naturally include the matching of the vocabulary. In that sense, the resulting query logs represent a rather worst-case scenario for evaluating the effectiveness of our indexing scheme.

Since we implicitly perform a vocabulary matching, we can first look at the resulting intersection of query terms and tags. As expected, the size of the intersection has further reduced, now comprising 5.17\% of all terms/tags in combination with \textsc{Delicious} and 4.44\% in combination with \textsc{Flickr}. With that, we now look more closely at the effects on the query log; see Table~\ref{tab:IntersectResultsQueryLog}. 

\begin{figure*}[htb]
\centering
\subfigure[queries frequency vs.  number of terms]{
\includegraphics[width=0.31\textwidth]{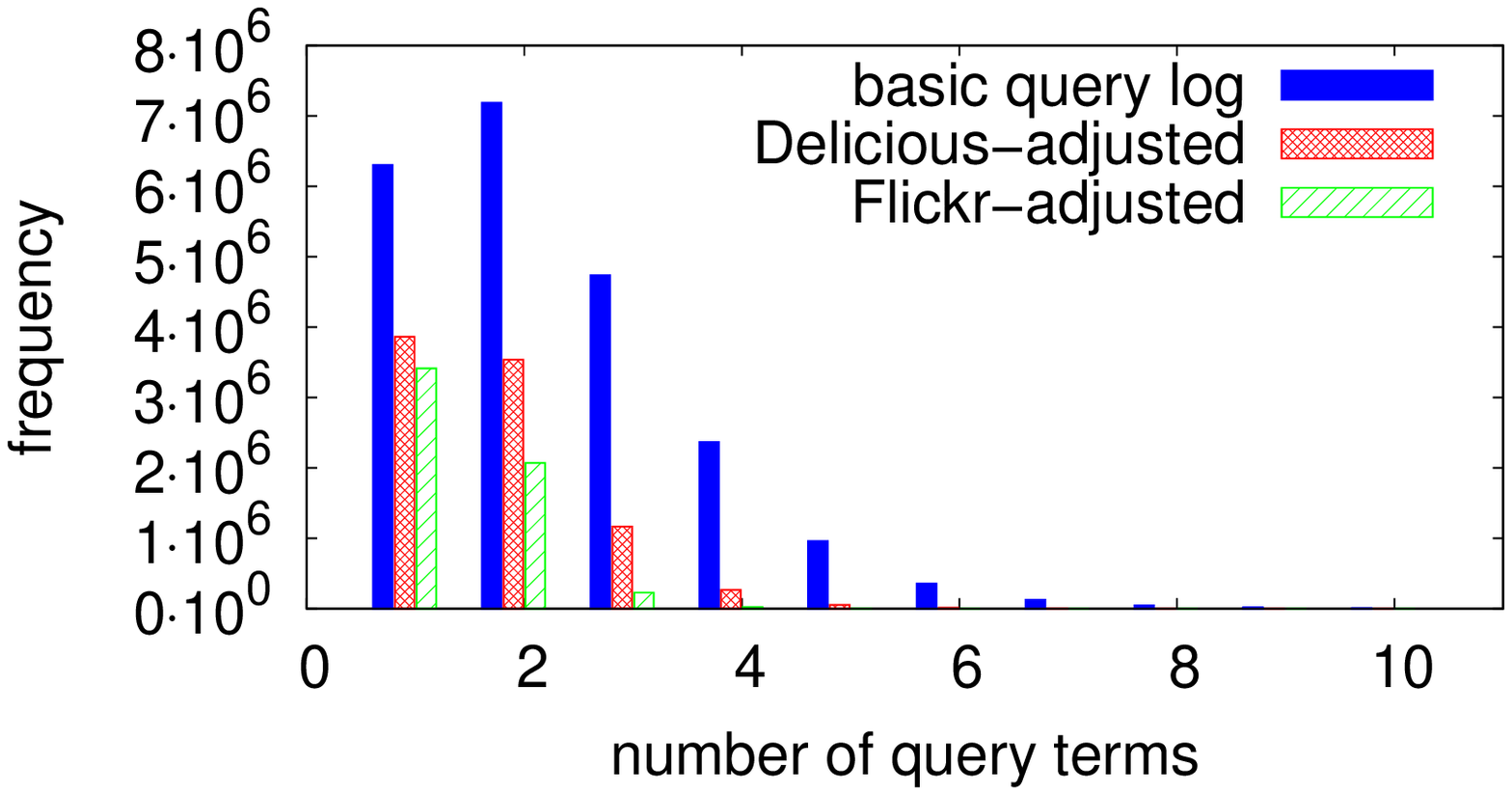}
\label{fig:aol_term_counts_adjusted}
}
\subfigure[key frequency distribution (\textsc{Delicious})]{
\includegraphics[width=0.31\textwidth]{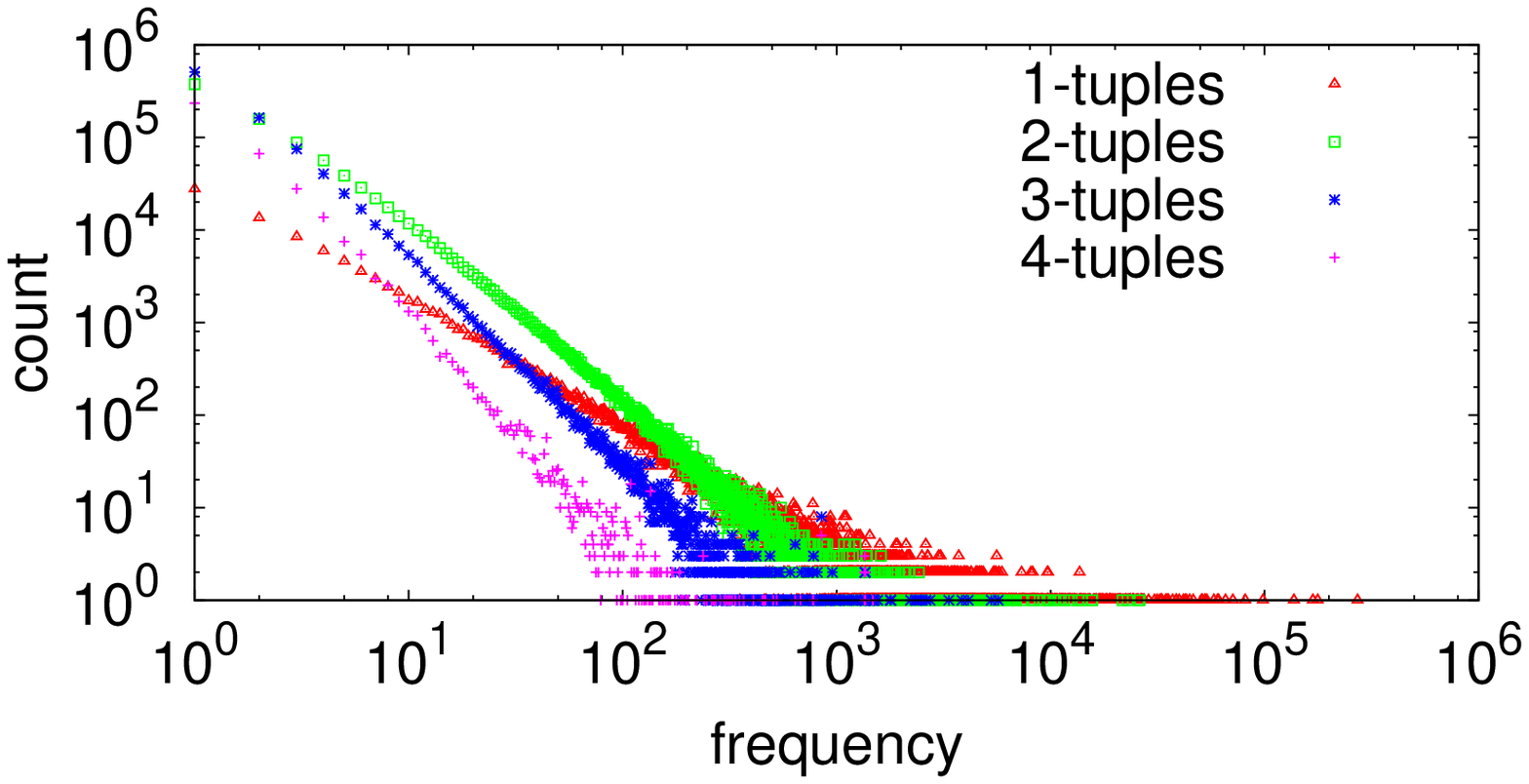}
\label{fig:aol_log_delicious}
}
\subfigure[key frequency distribution (\textsc{Flickr})]{
\includegraphics[width=0.31\textwidth]{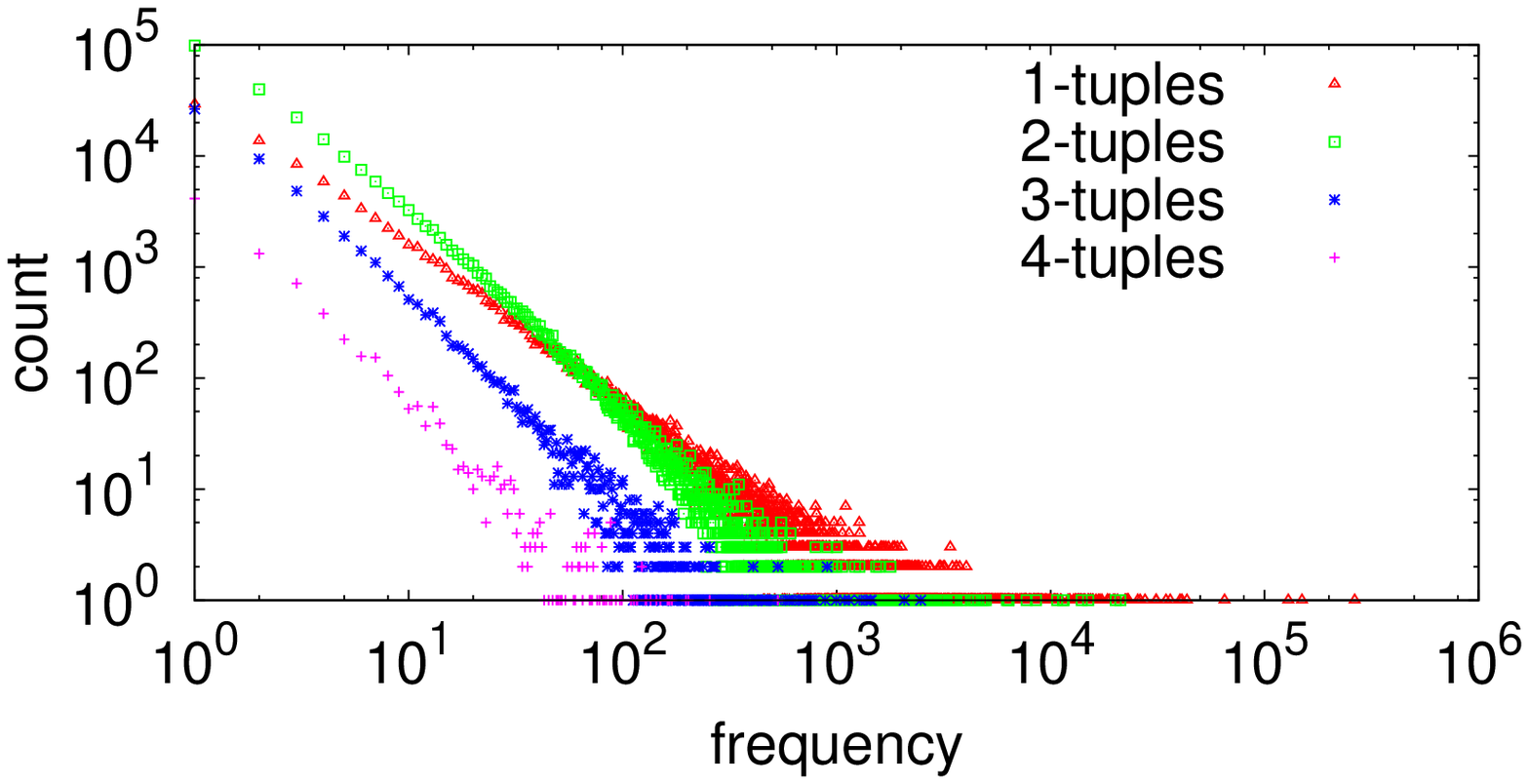}
\label{fig:aol_log_flickr}
}
\label{fig:adjusted_log_analysis}
\caption[]{Characteristics of query log-adjusted tag datasets.}
\end{figure*}

One can first observe that now the resulting dataset-adjusted query logs quite differ from each other. On all accounts, the query log related to the \textsc{Delicious} is closer to the basic query log compared to the one derived for the \textsc{Flickr}. Thus, with respect to the number of non-empty query results, the \textsc{AOL} query log and the \textsc{Delicious} tag dataset are closer to each other, than the query log to the \textsc{Flickr} dataset. However, the effect on the query logs when removing all queries yielding empty results is significantly more pronounced compared to matching the vocabulary alone. Here a larger number of both terms and eventually queries are no longer relevant. And also the average number of terms per query shows a drop from formerly 2.43 to 1.45 for the \textsc{Flickr} and to 1.78 for the \textsc{Delicious}-adjusted query log. To quantify in more detail how the size of queries shifted, Figure~\ref{fig:aol_term_counts_adjusted} shows the distribution of queries regarding their number of terms for both dataset-adjusted and the basic query log. In both adjusted datasets the queries with only one term dominate. Further, decline of the number of queries with large size is more pronounced in the adjusted datasets. Thus, ratio of queries returning empty results increase with the number of query terms. Both datasets-adjusted logs in comparison shows that this decline is more pronounced in the \textsc{delicious} log.

The previous results show that considering only queries returning a non-empty set of resource from the tag datasets clearly alters the basic characteristics of the query log. We therefore re-evaluate the relevant characteristics concerning our multi-tag indexing scheme. Figures~\ref{fig:aol_log_delicious} and ~\ref{fig:aol_log_flickr} show the relationships between the frequency of the keys (sets of terms) and the number of keys with a corresponding frequency for various key sizes for the \textsc{Delicious} and \textsc{Flickr} data set. Qualitatively, again the power-law relationship prevails for all sizes of keys. Table~\ref{tab:PowerLawAOLSpecific} shows the values for the scaling factor $alpha$ and the skew $\beta$ to fit the power law function $\alpha\cdot f^\beta$ where $f$ is the frequency with which a key occurred in all queries.
\begin{table}
	\centering
	\footnotesize
		\begin{tabular}{|l|c||c|c|c|c|}
			\hline
					\multicolumn{2}{|c||}{\multirow{2}{*}{$\alpha \cdot l^\beta$}} 				& \multicolumn{4}{c|}{size of key} \\
			\cline{3-6} 
					\multicolumn{2}{|c||}{}								& 1						& 2 				& 3 				& 4   \\
			\hline\hline
			\multirow{2}{*}{\textsc{Delicious}}		&	$\alpha$	& $2.8\cdot 10^4$			& $3.8\cdot 10^5$ 		& $5.1\cdot 10^5$	& $1.3\cdot 10^5$		\\
			\cline{2-6} 
						&	$\beta$ & $1.2$			& $1.4$ 		& $1.8$	& 	$1.9$		\\
			\hline \hline
			\multirow{2}{*}{\textsc{Flickr}}		&	$\alpha$	& $3.0\cdot 10^4$			& $1.0\cdot 10^5$ 		&  $2.7\cdot 10^4$	& $4.2\cdot 10^3$	\\
			\cline{2-6} 
						&	$\beta$ & $1.2$			& $1.4$ 		& $1.6$	& $1.7$		\\
			\hline
		\end{tabular}
	\caption{Power law parameters for dataset-adjusted query logs}
	\label{tab:PowerLawAOLSpecific}
\end{table}
Looking at the quantitative figures, the results reveal several things. Firstly, and also expectedly, since a large number of terms and queries have been removed, the absolute number of the scaling factor is significantly smaller for all keys in the adjusted query logs compared to the basic one. Secondly, the decreased average numbers of terms per query already indicates that the number of frequent keys drop more significantly for increasing key sizes than for the basic query log. And thirdly, the results also confirm the differences between the \textsc{Delicious} and \textsc{Flickr}-adjusted query log. Particularly the sharp increasing skew and decreasing scaling parameter for the \textsc{Flickr} log stand out.

Finally, we compared the absolute number of distinct keys derived from the adjusted query logs and the basic log; see Figure~\ref{fig:aol_term_set_distinct_sets} (cross- and left-hatched bars). Naturally, the adjusted query logs feature less distinct keys than the basic one. And further, one can clearly see the differences between both datasets with increasing key sizes. The larger the size of the keys the larger the differences between the number of distinct keys, where the \textsc{Delicious}-adjusted query log is much closer to the basic one. Particularly for keys of size 3 and 4 the number of keys is only just a fraction compared to the numbers that can be derived from the basic query log.
\\
\\
\textbf{Assumptions and evaluation method.}
We assume a distributed key-value store for managing the inverted index. In this evaluation, we ignore node failures. Particularly for single-term keys, we assume that they are always available. Since we do not consider locality-preserving data placement strategies etc., we assume the worst case, i.e. a sufficiently large number of back end nodes so that all relevant keys for processing a query or for propagating an update reside on different nodes. In our experiments we measure three parameters to evaluate the overall system performance:

\textit{Number of contacted keys (CK).} Parameter $CK$ represents all single accesses to keys in the inverted index, both read and write accesses.

\textit{Number of invoked keys (IK).} As subset of $CK$, the $IK$ is the number keys whose inverted list is read while performing queries, updates or resuming keys. 

\textit{Number of transferred resources (TR).} The most relevant parameter to describe the performance is $TR$ representing the number of resources that are actually transferred for processing queries, updates and resuming keys.

\textit{Number of handled resources (HR).} To investigate the effect of caching on the shifting of the load between gateway and back end nodes we quantify the load of nodes by counting the numbers of resources they handle, i.e. the resources nodes read from or write to secondary storage and send or receive via the network interface.

With these parameters and our assumption of a sufficiently large number of back nodes, our results are independent from the actual number of back end nodes in the systems. In other words, adding further nodes would have no impact on the results.

\subsection{Multi-Term Indexing}
\label{subsec:eval_mtk}
\begin{figure*}
\parbox{.30\linewidth}{
	\centering
		\includegraphics[width=0.32\textwidth]{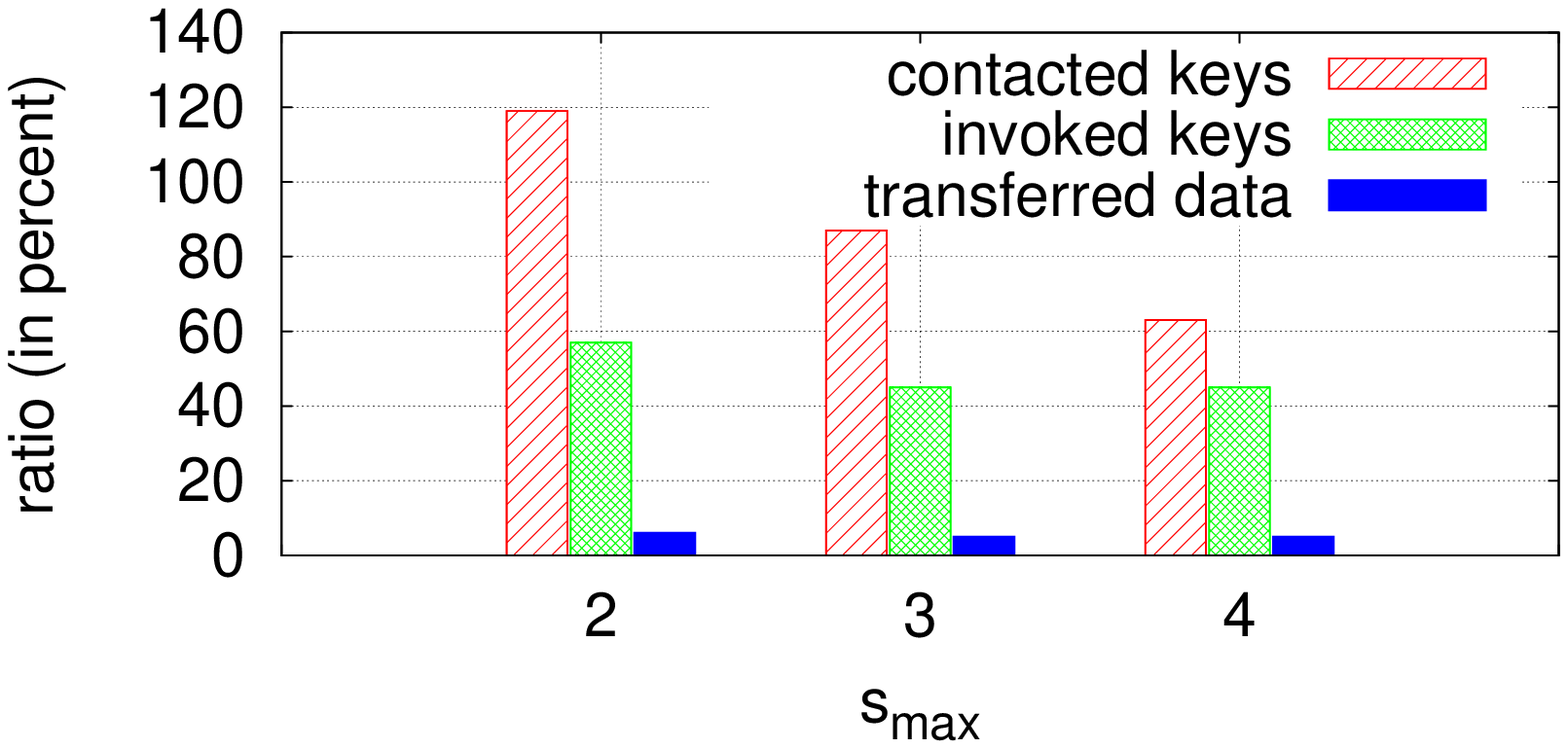}
	\caption{STK vs MTK (best case) for \textsc{Delicious} tag data set}
	\label{fig:best_vs_worst_delicious}
}
\hfill
\parbox{.30\linewidth}{
	\centering
		\includegraphics[width=0.32\textwidth]{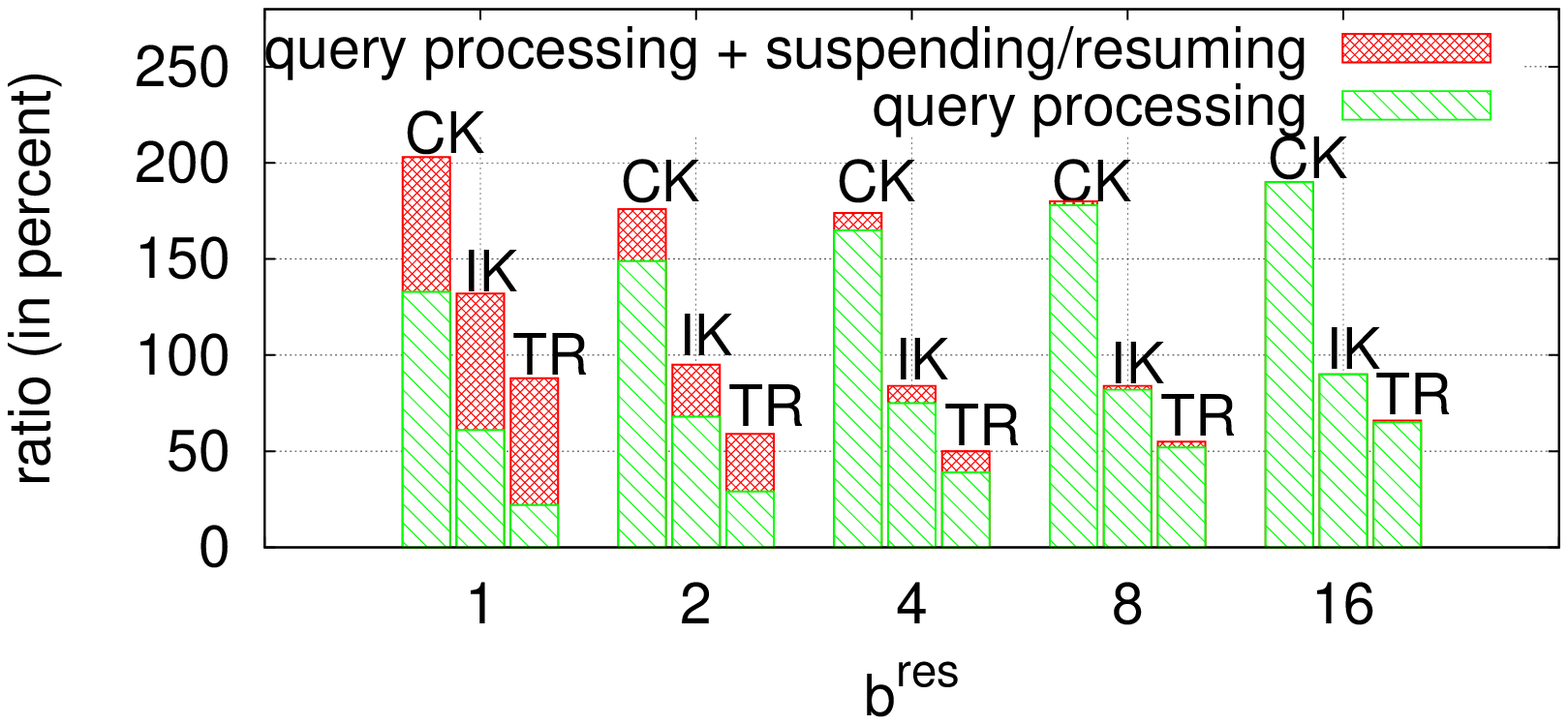}
	\caption{Varying minimum number of set bits $b^{res}$}
	\label{fig:resumingBres}
}
\hfill
\parbox{.30\linewidth}{
	\centering
		\includegraphics[width=0.32\textwidth]{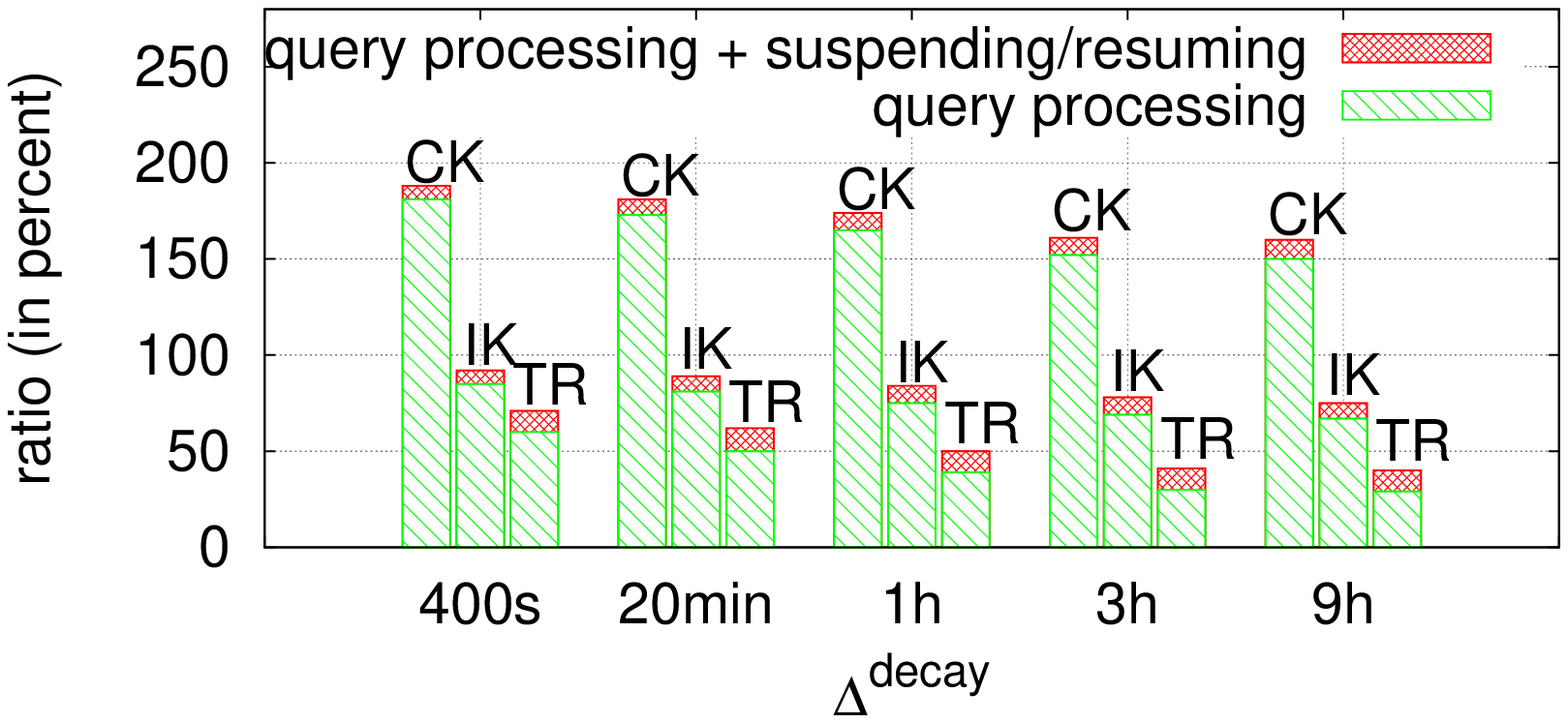}
	\caption{Varying time $\Delta^{decay}$ for periodic shifting}
	\label{fig:resumingDelta}
}
\end{figure*}

We evaluate our approach using multi-term keys, henceforth denoted by MTK, against the na\"{i}ve one based solely on single-term keys (STK). To make the results comparable between each other, we compute the relative differences between our MTK and STK, where we normalize the load for the STK to 100\%. Since the processing of single-term queries is identical for STK and MTK, we use only queries with more than one term throughout our experiments. We performed all experiments on both the \textsc{Delicious} and \textsc{Flickr} data set, using the corresponding adjusted query logs. While the absolute figures may vary, the quantitative results are very similar for both data sets. Therefore, due to space constraints, we present only the results for the \textsc{Delicious} data set.
\\
\\
\textbf{STK vs. MTK (best case):} We first compare STK and MTK with all multi-term keys that are relevant for answering a query being available in the index. This case is of theoretical nature, since it requires the index to anticipate all relevant keys a-priori. Further, we consider no updates in this experiment. Comparing both cases allows estimating a (theoretical) upper bound for the improvement with the usage of multi-term keys. Figure~\ref{fig:best_vs_worst_delicious} shows the result for \textsc{Delicious} and various values of $s_{max}$ ($t_{max}=20$).

Since more and larger keys are available, MTK performs better for increasing values of $s_{max}$ (not visible for $TR$). However, the improvements quickly converge, since less and less queries benefit from larger keys. Although for MTK the number of contacted keys $CK$ is in $O(|q|^{s_{max}})$ -- compared to $O(|q|)$ for STK -- the result for CK are in most cases still better when using multi-term keys. This is due to the fact that in a perfect index, each query with $\leq s_{max}$ query terms can be answered by contacting only the corresponding key. Only for $s_{max}=2$, $CK$ is higher for MTK, since the query log contains too much queries $q$ with $|q| > s_{max}$. The most important result concerns the differences between the number of transferred resources $TR$. Given a perfect multi-term index, only about 5\% of resources are transferred during query processing, compared to STK. Thus, given our tag data set and query log, this represents the best case we can achieve. For the rest of our evaluation we set $s_{max}=3$, representing the most practical value.
\\
\\
\textbf{Resuming keys.} We now consider the suspending and resuming of keys depending on their popularity derived from a query history. While suspending keys is bandwidth-neutral, resuming keys add to the workload for processing user queries. Our mechanism to measure the popularity of a key features four parameter: $\ell$, $\Delta^{decay}$, $b^{res}$ and $b^{susp}$ (cf. Section~\ref{subsec:indexSuspendingResuming}). Again, we compare our multi-term key approach against the na\"{i}ve one based on single-term keys.

In the first test we vary the minimum of set bits in a bit vector $\mathcal{B}_k$ specifying when to resume key $k$. We set $b^{susp}=0$, i.e., we suspend keys when no bit is set in the corresponding bit vector. Further, we set $\ell = 24$ and $\Delta^{decay}=1\mathrm{h}$. Thus, each request on a key $k$ is represented as a set bit in $\mathcal{B}_k$ for 24h. Figure~\ref{fig:resumingBres} shows the results for $b^{res}\in \{1,2,4,8,16\}$. In this figure we differentiate between the load only induced by processing user queries and the overall load to emphasize on the additional load caused by resuming keys. Processing user queries clearly benefits from smaller values for $b^{res}$, since the number of available multi-term keys increases, see Table~\ref{tab:resumingIndexSizes}. However, the frequent resuming of keys adds to the overall load. For increasing values for $b^{res}$, since less keys are available in the index, the ratio between the load for resuming keys and processing queries shifts toward a higher load for processing user queries, while the overall load stays quite equal. If $b^{res}$ becomes too large, and therefore the number of available keys to small, the decreasing load for resuming keys can no longer compensate for the increasing load caused by user queries, and the overall load increases.

\begin{table}[htb]
	\centering
	\footnotesize
		\begin{tabular}{|c|c|c|c|c|}
											\multicolumn{5}{l}{\textbf{Resuming keys: various values for $b^{res}$}} \\
			\hline
					1&										 2						& 4 				& 8 				& 16   \\
			\hline\hline
					3.08\%&								1.38\%		 						& 0.53\% 				& 0.12\% 				&  0.0\%$^\ast$  \\
			\hline
			\end{tabular}\\
			\vspace{1em}
			\begin{tabular}{|c|c|c|c|c|}
											\multicolumn{5}{l}{\textbf{Resuming keys: various values for $\Delta^{decay}$}} \\
			\hline
						400s	& 20min	& 1h	& 3h & 9h   \\
			\hline\hline
					0.02\%&								0.07\%		 						& 0.53\% 				& 1.43\% 				&  1.5\%  \\
			\hline
		\end{tabular}

	\caption{Relative index size compared to optimal index with all relevant keys available ($^\ast$practically empty)}
	\label{tab:resumingIndexSizes}
\end{table}

In a second test we modify $\Delta^{decay}$, i.e. the time span a request on a key $k$ is represented by set bit in $\mathcal{B}_k$. Again, $\ell\!=\!24$ and $b^{susp}\!=\!0$. In this test, we set $b^{res}\!=\!4$. Thus, the results for $\Delta^{decay}\!=\!1\mathrm{h}$ is the same as in Figure~\ref{fig:resumingBres} for $b^{res}\!=\!4$. Figure~\ref{fig:resumingDelta} shows the results for $\Delta^{decay}\in \{400\mathrm{s}, 20\mathrm{min}, 1\mathrm{h}, 3\mathrm{h}, 9\mathrm{h}\}$, and Table~\ref{tab:resumingIndexSizes} the resulting index sizes. Here, the load for resuming keys hardly changes for different values of $\Delta^{decay}$, since $\Delta^{decay}$ only specifies how $long$ a key is kept in the index and not how $soon$. The overall performance increases for increasing values for $\Delta^{decay}$, since more and more keys are kept in the index, see Table~\ref{tab:resumingIndexSizes}. Thus, since the number of multi-term keys are with respect to the storage requirements are still reasonable low, larger values for $\Delta^{decay}$ are beneficial. However, the more multi-term keys are available in the inverted index the higher the expected overhead to update them.
\\
\\
\textbf{Handling updates.} Our proposed update mechanism propagates changes on the tag data only to the corresponding single-term key. As a consequence, processing queries solely based on single-term keys (STK) and exploiting available multi-term keys might yield different results. To quantify this, we compared the results for both approaches on the inverted index, after various numbers of updates on the inverted lists of single-term keys. We assumed an optimal index, i.e. all relevant multi-term keys are available. Regarding updates, this is the worst-case scenario, since MTK never has to invoke up-to-date single-term keys. Table~\ref{tab:averageOverlap} shows the results. Naturally, for an increasing number of updates, the average overlap between query results decreases. Which degree of deviation is acceptable is a system design decision.
\begin{table}[htb]
	\centering
		\footnotesize
		\begin{tabular}{|c|c|c|c|c|c|c|}
			\hline
										&	\multicolumn{5}{p{5cm}|}{\textbf{Changes in the inverted lists of single-term keys}} \\
			\hline
				& 0.25\% & 0.5\%	& 1\%	&										 2\%						& 4\% 		\\
			\hline\hline
				\textbf{overlap}	& 99.1\% &  98.6\%&								97.6\%		 						& 95.7\% 				& 92.3\% 	\\	
			\hline
			\end{tabular}
			\caption{Average overlap of query results between na\"{i}ve and multi-term approach for various rates of updates}
	\label{tab:averageOverlap}
\end{table}

For our subsequent experiments we make the following assumptions: Users perform 150 actions per minute, which is more than twice the figure we derived from the \textsc{Delicious} data set (cf.~Table~\ref{tab:BasicNumbersOfDatasets}). Further, we aim to ensure an overlap of above 99\%. Thus, we only tolerate 0.25\% of changes in the inverted lists of single-term keys. With that, given the number of $\sim$10.9 million inverted list entries of single-term keys, we have to update all available multi-term keys at least every $\Delta^{update} = 3\mathrm{h}$. Again, we vary $\Delta^{decay}$ and keep the other parameters fix ($\ell\!=\!24$, $s_{max}\!=3\!$, $b^{res}\!=\!4$, $b^{susp}\!=\!0$).

\begin{figure}
	\centering
		\includegraphics[width=0.45\textwidth]{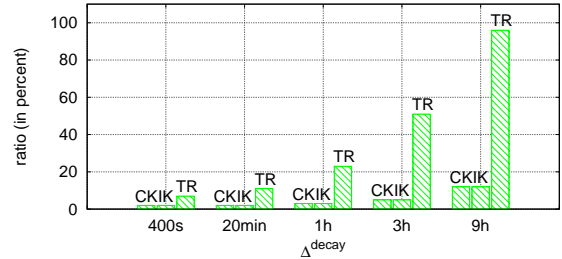}
	\caption{Direct propagation of updates to all relevant keys vs. only single-term key updates with incremental updates of multi-term keys}
	\label{fig:all_vs_single-key}
\end{figure}

We first compared both alternatives for handling updates, the direct propagation of all keys derived from a query and the propagation of updates only to the corresponding single-term keys in combination of incremental updates for multi-term keys; see Figure~\ref{fig:all_vs_single-key}. The load for the propagation of updates to all single- and multi-term keys does not depend on the current state of the inverted index. Since the load for incremental updates increases for larger numbers of available multi-term keys in the index, the performance gain due to incremental updates decreases for larger values of $\Delta^{decay}$. Thus, for very large values of $\Delta^{decay}$, a propagation to all keys will eventually outperform the approach of incremental updates, particularly regarding the number of transferred resources.

\begin{figure}
	\centering
		\includegraphics[width=0.45\textwidth]{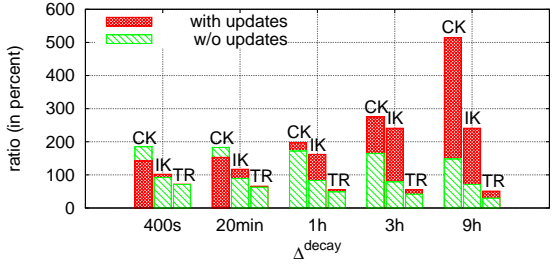}
	\caption{STK vs. MTK with and without updates}
	\label{fig:naive_vs_optimized}
\end{figure}

Finally, we evaluated the overall system performance in the presence of updates, again comparing STK and MTK. STK only requires the propagation of updates to single-term keys; MTK additionally requires incremental updates. The parameter settings are the same as in previous test. Figure~\ref{fig:naive_vs_optimized} shows the result. Since the incremental update process of a multi-term key contacts each corresponding single-term key, the number of contacted keys significantly increases for larger values of $\Delta^{decay}$ (= larger number of available keys). Further, now in the presence of updates, also the saved number of transferred resources due to MTK does no longer benefit from many available key.
\\
\\
Summing up, our results clearly indicate the trade-off between the query processing performance and the load for maintaining the index in the presence of updates with respect to the number of available multi-term keys in the index. A large index speeds up the evaluation of queries, but causes high maintenance costs, and vice versa. Although MTK involves increasing costs for the index maintenance, the improvements regarding the overall bandwidth consumption significantly outweighs the maintenance costs. Despite our worst cases assumptions for the parameter settings, MTK reduces the number of transferred resources to less the 50\% compared to STK. In real-world systems, we expect even better results. 

\subsection{Caching}
\label{subsec:eval_caching}
We now investigate the effect of caching on both the single-term and multi-term indexing. Henceforth, $\mathrm{STK_C}$ denotes STK with additional caching, and $\mathrm{MTK_C}$ denotes MTK with caching. We consider uniform caching, i.e. caching each key on all gateway nodes, as well as dedicated caching, i.e. caching each key on a single gateway node. 
For the distributed index we keep the parameter settings from previous experiments. To be more specific, $\Delta^{update} = 3\mathrm{h}$, $\Delta^{decay}=1\mathrm{h}$, $\ell\!=\!24$, $s_{max}\!=3\!$, $b^{res}\!=\!4$, $b^{susp}\!=\!0$. 
If not stated otherwise, we set $c^{del} = 0$, and vary $c^{ins}$, i.e. the minimum number of set bits in $\mathcal{B}_k$ to cache key $k$. Since $b^{res}\!=\!4$ and $\ell\!=\!24$, and since we stores only keys available in the index, $4 \leq c^{ins} \leq 24$. Further, we assume an architecture with 5 gateway nodes.

\begin{figure*}
\parbox{.30\linewidth}{
	\centering
		\includegraphics[width=0.32\textwidth]{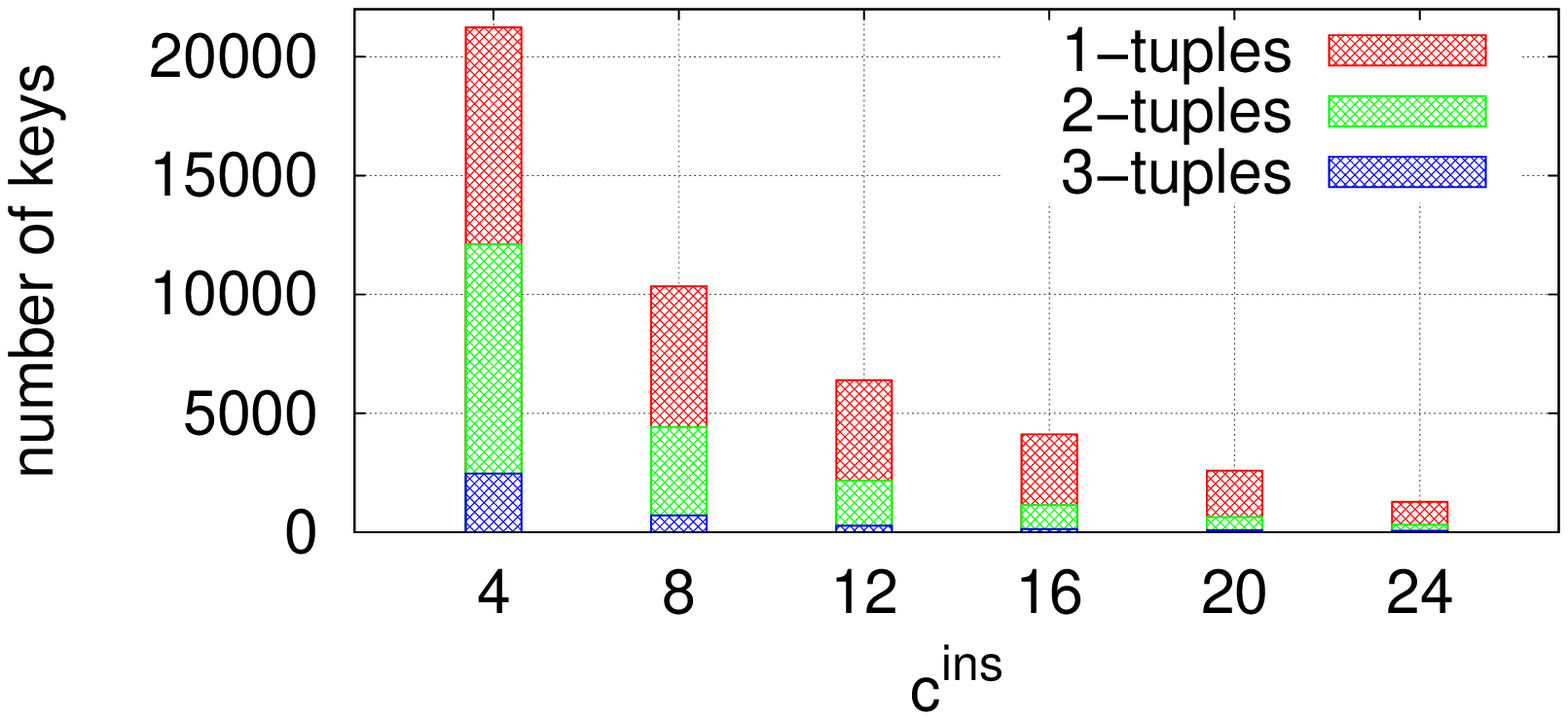}
	\caption{Number of single-term and multi-term keys in the cache}
	\label{fig:key_ratio_c}
}
\hfill
\parbox{.30\linewidth}{
	\centering
		\includegraphics[width=0.32\textwidth]{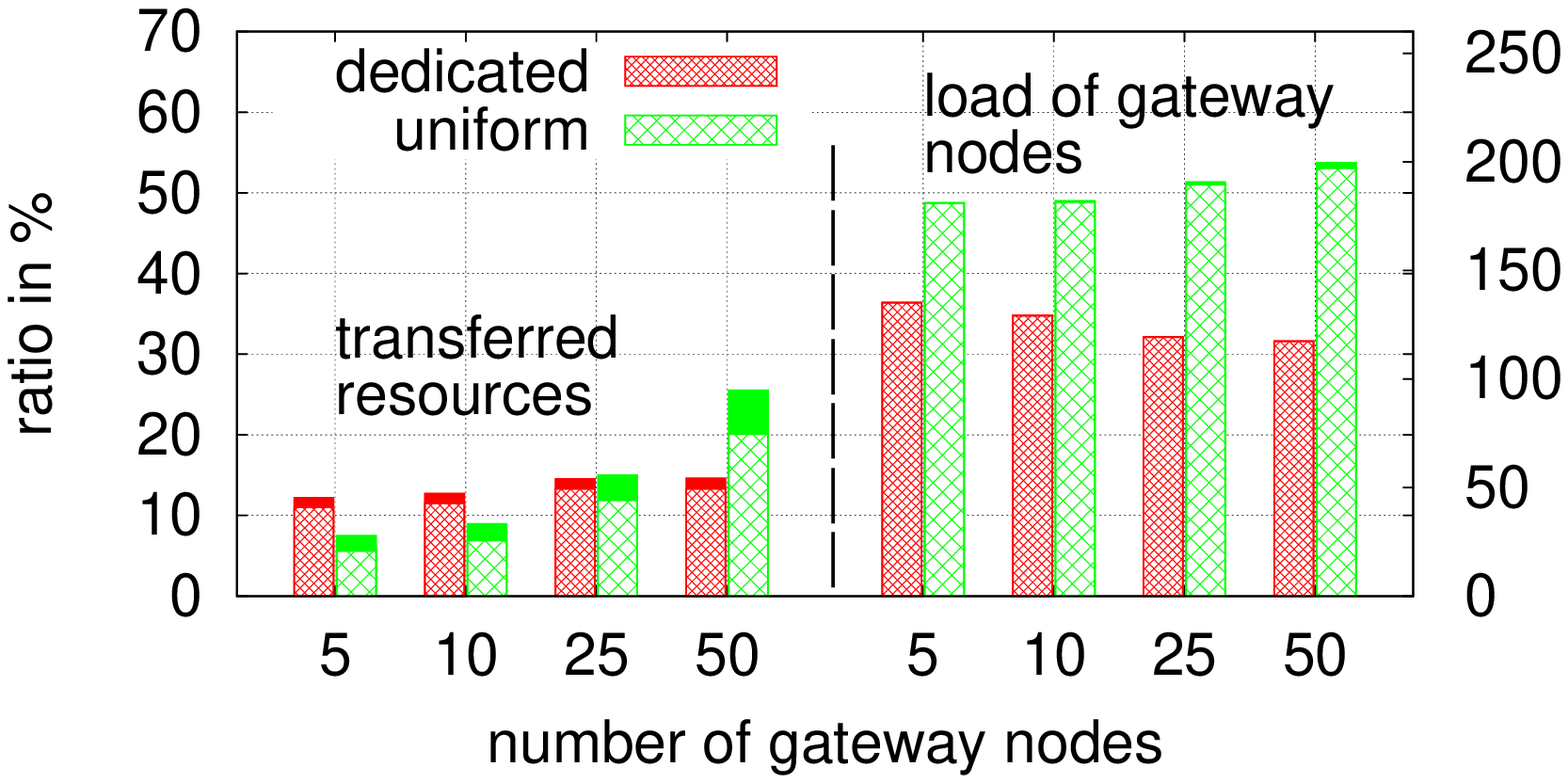}
	\caption{Effect of the number of gateway nodes}
	\label{fig:MDK_vs_MDKc_NumOfGN}
}
\hfill
\parbox{.30\linewidth}{
	\centering
		\includegraphics[width=0.32\textwidth]{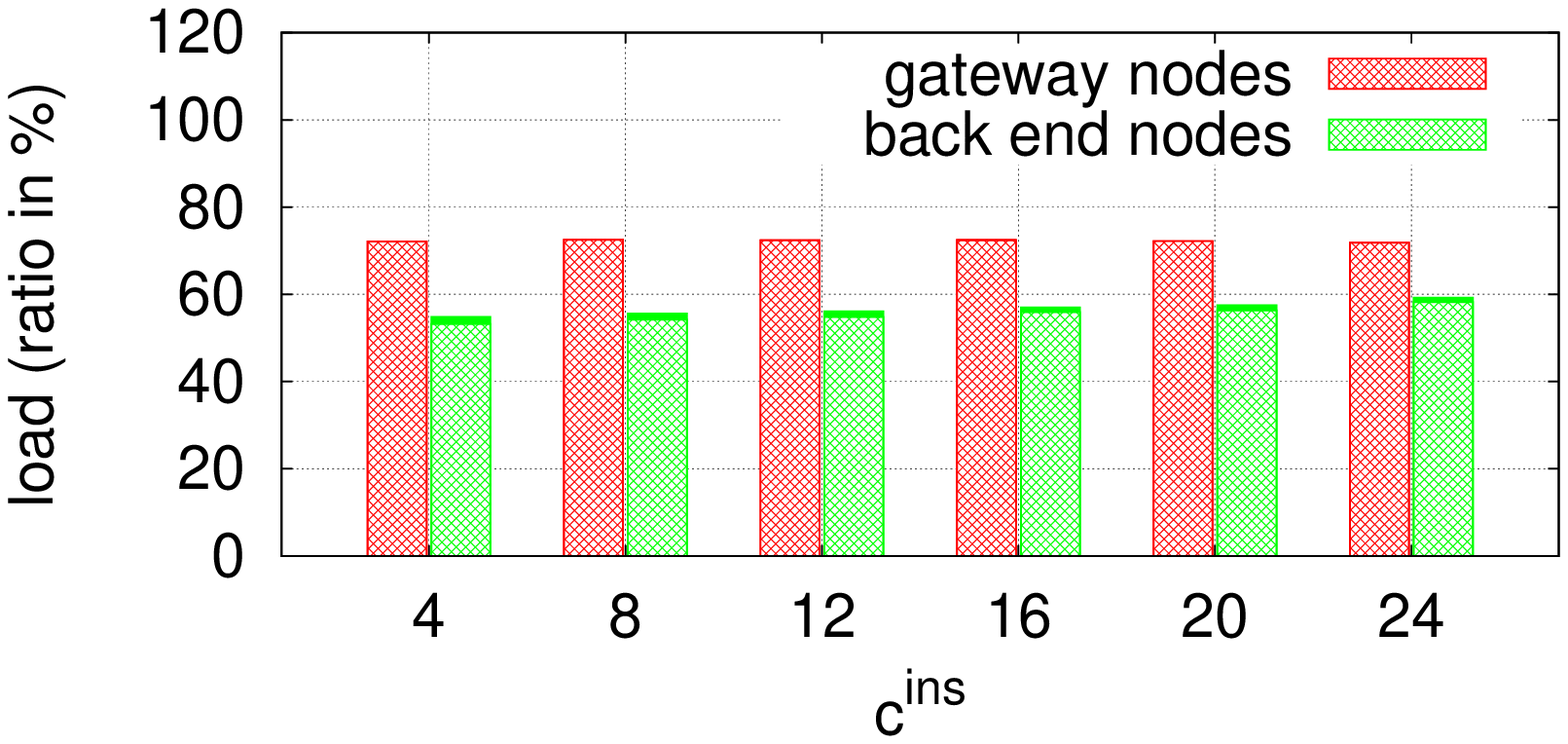}
	\caption{$\mathrm{STK_C\ vs.\ MTK_C}$: Effect of caching on the load of nodes}
	\label{fig:SDKc_vs_MDKc_scaled}
}
\end{figure*}

To quantify the performance by means of network traffic we measure the number of transferred resources with then back end.
Additionally, we measure the load of gateway nodes (GN) and back end nodes (BN) by means of handled resources to see how caching shifts the load of back end nodes to the gateway nodes. Similar to previous experiments, to highlight the impact of updates we all figures show the results with and without the shares attributed to updates (fully filled part of bars in each following figure). In our experiments we evaluate the relative differences between the following settings: $\mathrm{STK\ vs.\ STK_C}$ and $\mathrm{MTK\ vs.\ MTK_C}$, quantify effect of caching based on a distributed without and with the support of multi-term keys, and $\mathrm{STK_C\ vs.\ MTK_C}$ 
%
%
%
\begin{figure*}[htb]
\centering
\subfigure[$\mathrm{STK\ vs.\ STK_C}$]{
\includegraphics[width=0.31\textwidth]{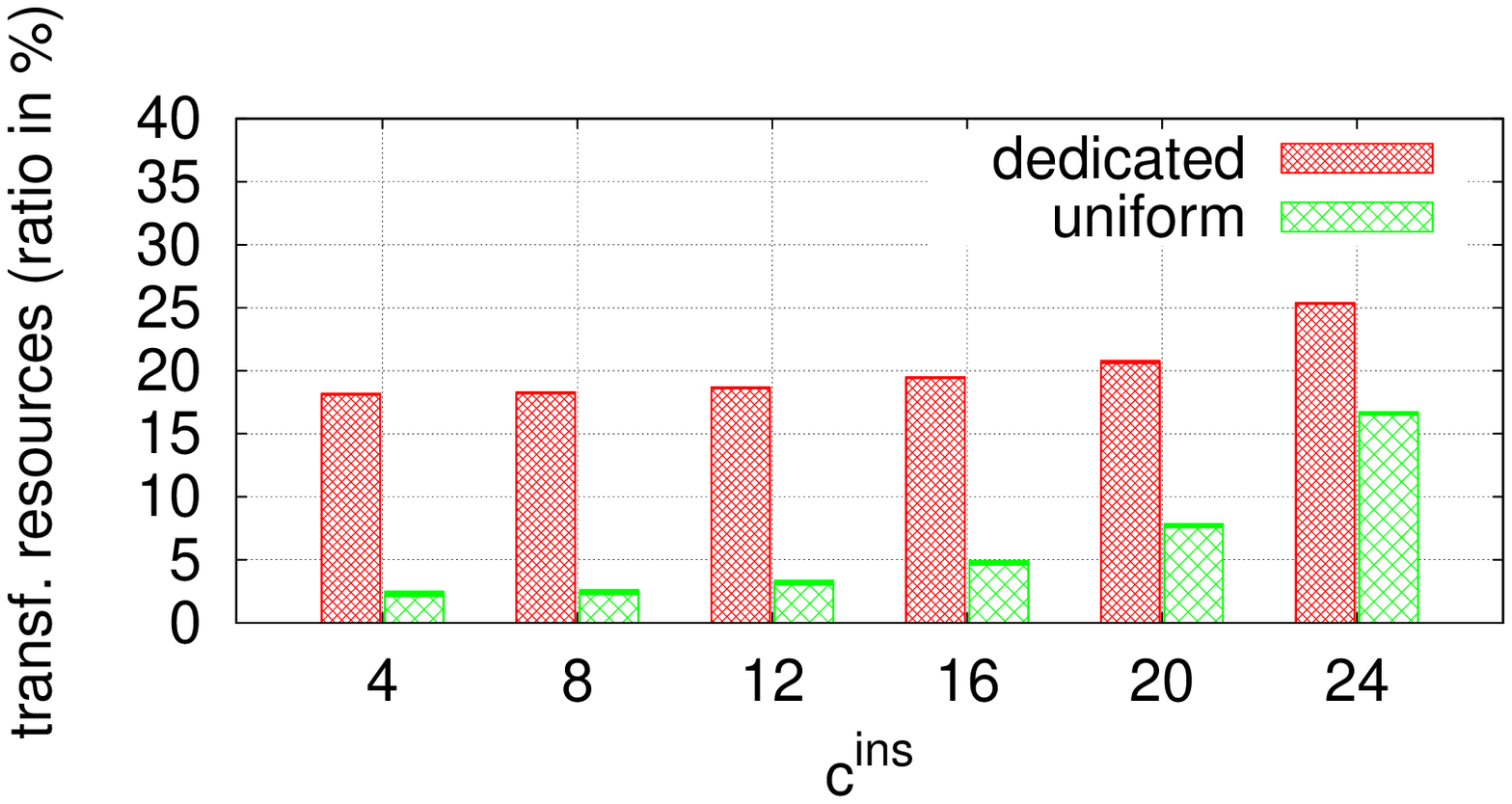}
\label{fig:trans_SDK}
}
\subfigure[$\mathrm{MTK\ vs.\ MTK_C}$]{
\includegraphics[width=0.31\textwidth]{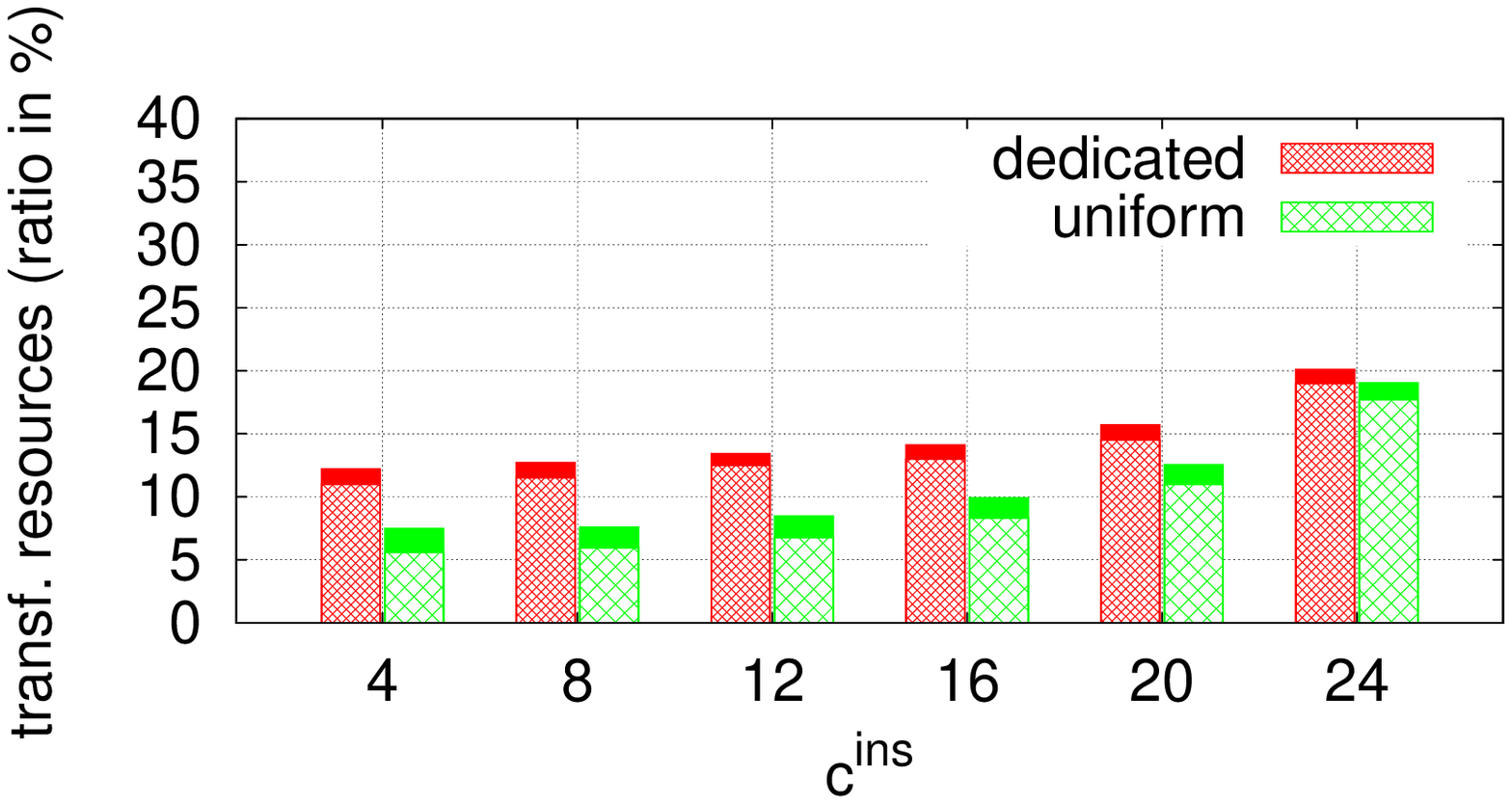}
\label{fig:trans_MDK}
}
\subfigure[$\mathrm{STK_C\ vs.\ MTK_C}$]{
\includegraphics[width=0.31\textwidth]{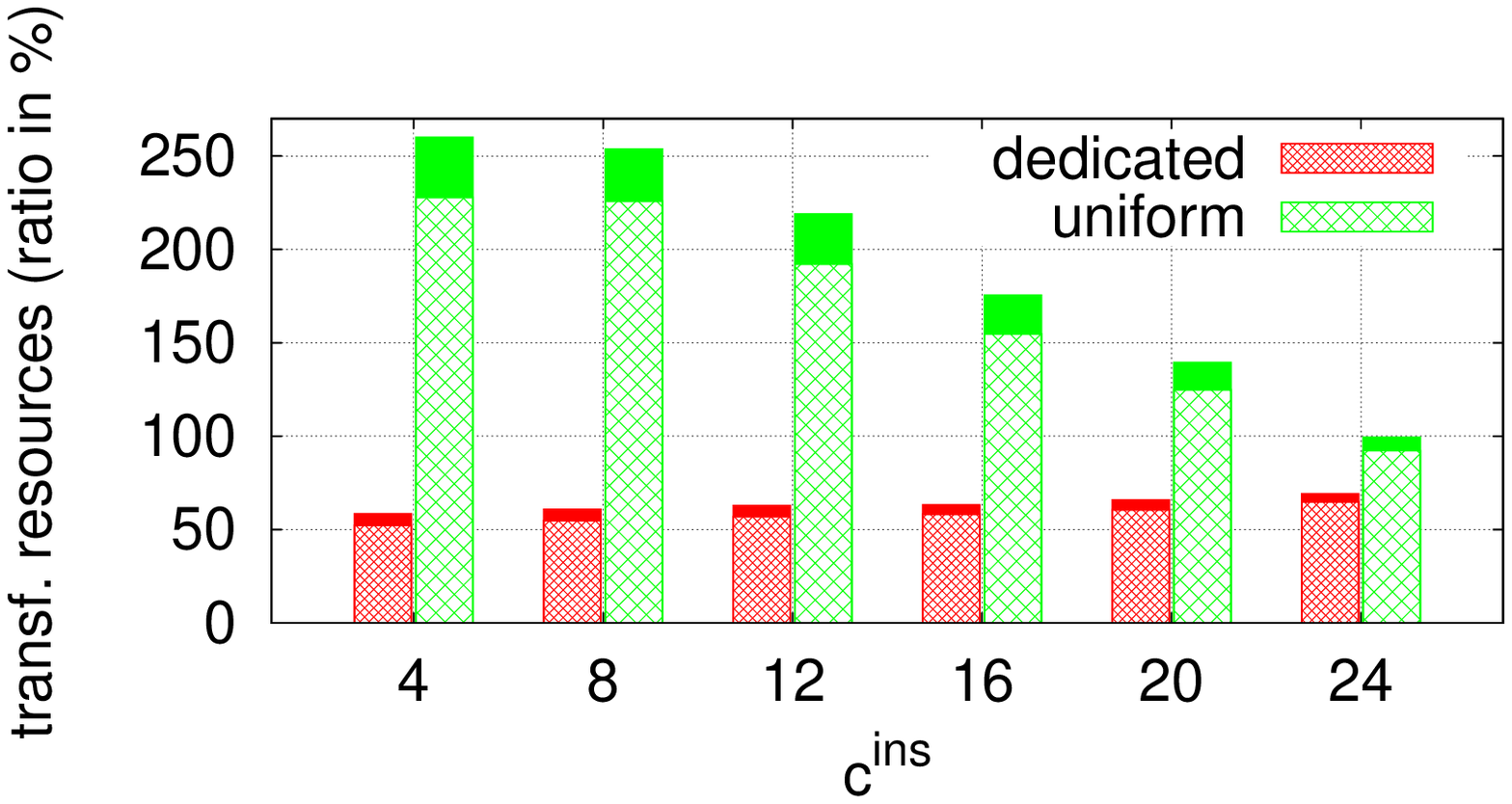}
\label{fig:trans_STK_MTK}
}\\
\subfigure[$\mathrm{STK\ vs.\ STK_C}$]{
\includegraphics[width=0.31\textwidth]{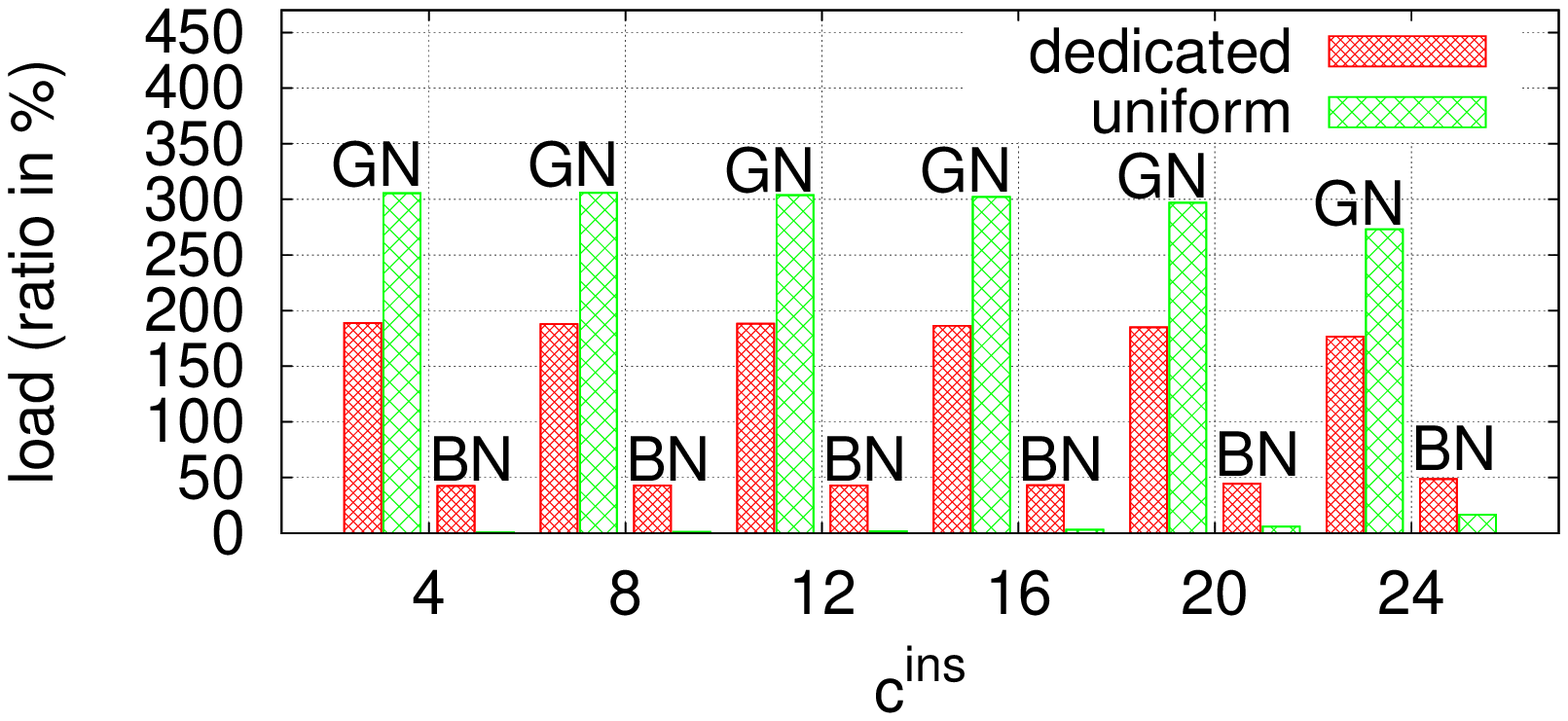}
\label{fig:load_SDK}
}
\subfigure[$\mathrm{MTK\ vs.\ MTK_C}$]{
\includegraphics[width=0.31\textwidth]{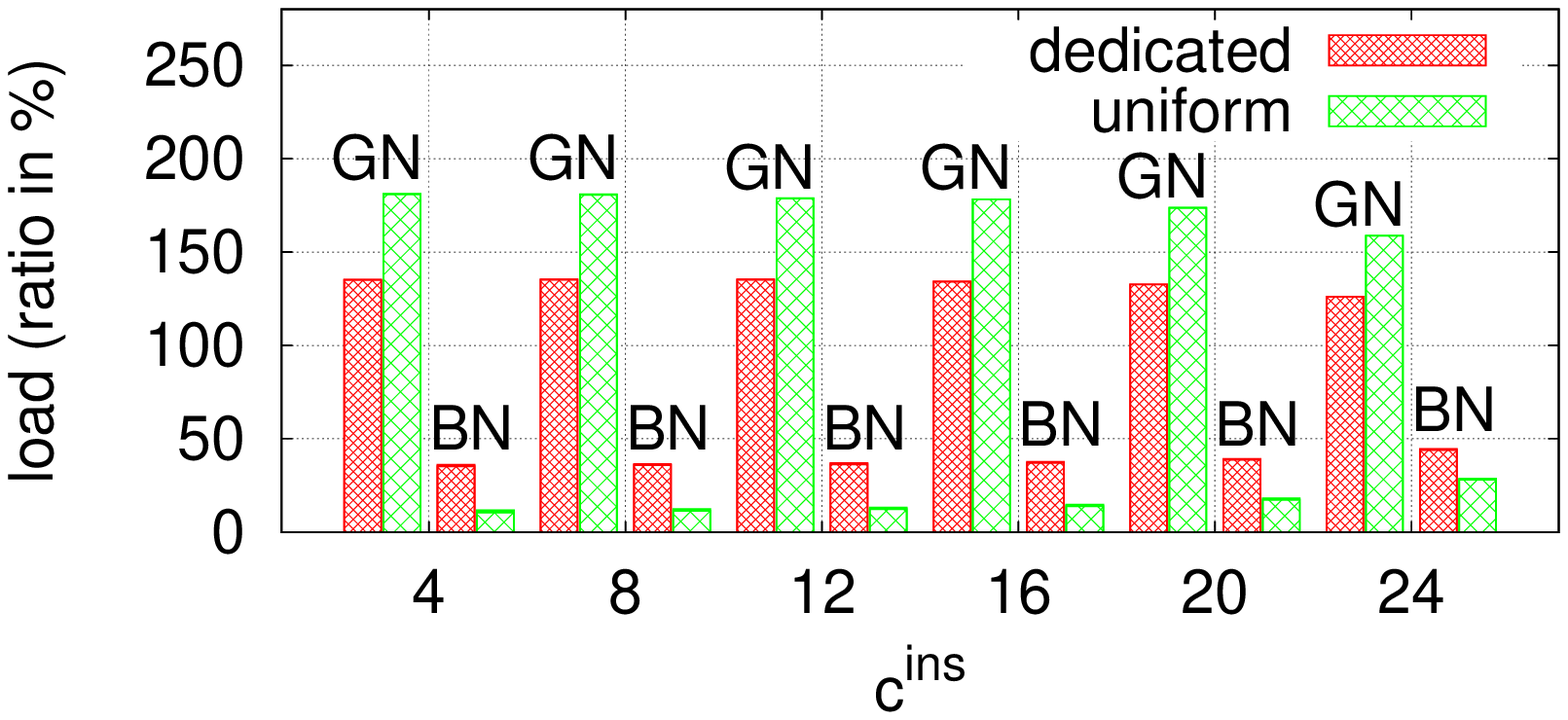}
\label{fig:load_MDK}
}
\subfigure[$\mathrm{STK_C\ vs.\ MTK_C}$]{
\includegraphics[width=0.31\textwidth]{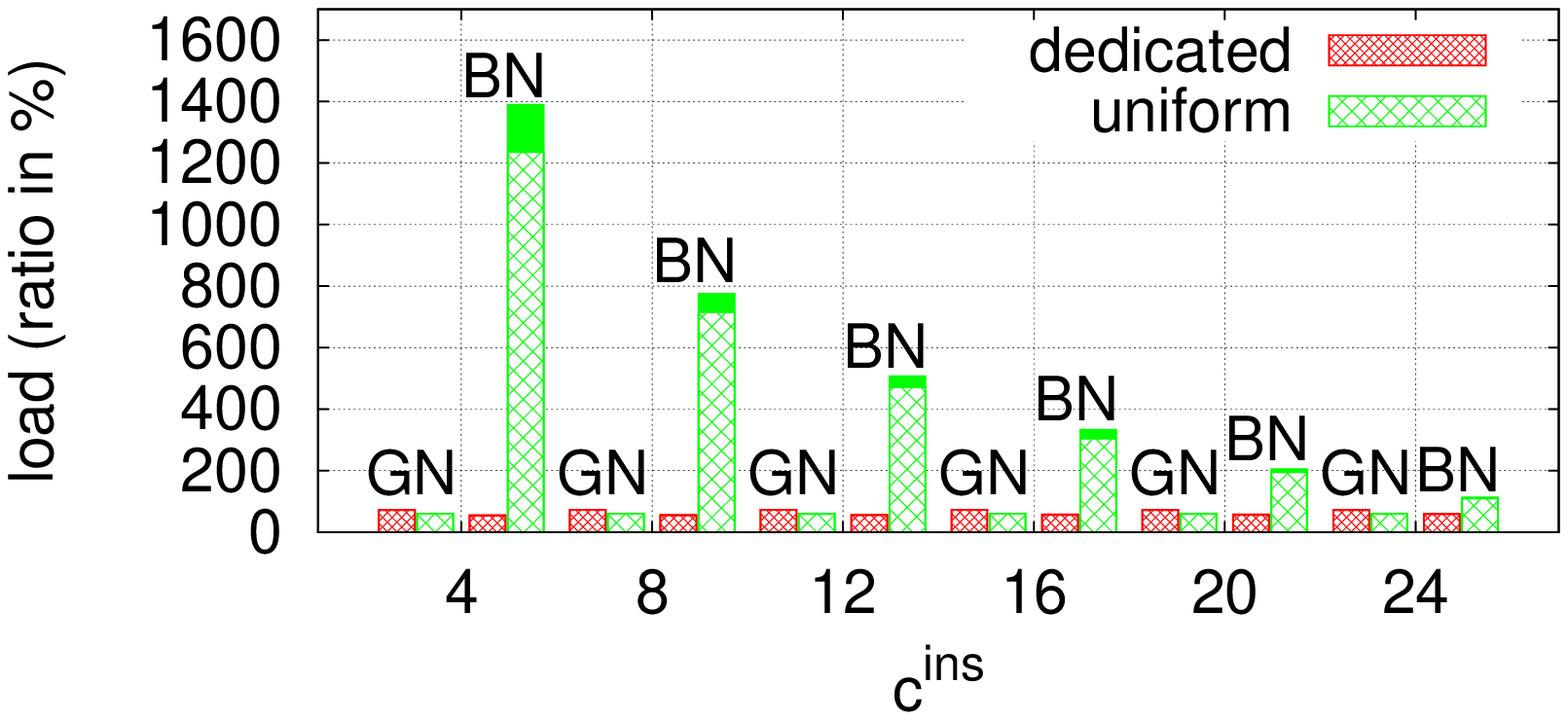}
\label{fig:load_STK_MTK}
}
\caption[]{Effect of caching on the number of transferred resources and the load of gateway and back end nodes.}
\label{fig:caching}
\end{figure*}
\\
\\
\textbf{Impact of updates.} All results indicate that the impact of updates is negligible. This has two reasons. Firstly, since the corresponding single-term keys of a popular multi-term key are also popular, most cached keys are single-term keys. Compared to processing queries, the number of transferred and handled resources for handling single-key updates is very small. And secondly, an incremental update of a multi-term key only needs to be propagated from the index to the cache if the update yielded a non-empty result. And in case of a non-empty-result, the result size tends to be very small. Thus, for the performance of the cache, updates are minor issue, even when supporting multi-term keys in the distributed index.
\\
\\
\textbf{General effects of caching.} Regarding the number of transferred resources both STK, Figure~\ref{fig:trans_SDK}, and MTK, Figure~\ref{fig:trans_MDK}, benefit significantly from caching. Naturally, caching shifts the overall load from the back end nodes to the gateway nodes, Figures~\ref{fig:load_SDK} and~\ref{fig:load_MDK}. The decrease in the number of transferred resources and the changes in the load of nodes depend on the cache size. The more keys are in the cache, the smaller the network overhead but the higher the load of the gateway nodes, and vice versa. Since only multi-term keys that are available in the index are eligible for caching, a large set of parameters affect the number of cache entries. This includes the parameters of the inverted index ($\Delta^{decay}$, $\ell$, $b^{res}4$, $b^{susp}$) as well as the parameters for the cache maintenance  ($c^{ins}$, $c^{del}$). In the experiments presented, to ease clarification, we only varied $c^{ins}$. Although the number of cached keys vary significantly for different values of $c^{ins}$, see Figure~\ref{fig:key_ratio_c}, even for the maximum value, $c^{ins}=\ell$, the cache still contains the most popular keys so that the results for different values of $c^{ins}$ differ not very pronounced.
\\
\\
\textbf{Uniform vs. dedicated caching.} 
In terms of cache hits, uniform caching represents the optimal case since each gateway node stores the complete cache. In case of dedicated caching, a gateway node handling a multi-term query $q$ does not benefit from keys that are both cached and relevant to answer $q$ but cached on other gateway nodes. As a result, uniform caching reduces the number of transferred resources more than dedicated caching. This holds for both STK, Figure~\ref{fig:trans_SDK}, and MTK, see Figure~\ref{fig:trans_MDK}. The difference between the results for uniform and dedicated caching decreases for higher values of $c^{ins}$, i.e. smaller cache sizes. This due to the fact that for larger values of $c^{ins}$ the ratio of single-term keys increases, see Figure~\ref{fig:key_ratio_c}. Single-term queries are always forwarded to the gateway node that potentially stores the corresponding key. Thus, in case of single-term queries uniform and dedicated caching yield the same results. 
The results for the load of gateway nodes is in line with results for the number of transferred resources, see Figures~\ref{fig:load_SDK} and ~\ref{fig:load_MDK}. Compared to dedicated caching, uniform caching involves much more overhead to cache popular keys. However, due to the higher number of cache hits, the load of the back end nodes is lower than for dedicated caching. Again, for larger values of $c^{ins}$, thus for a higher ratio of single-term keys in the cache, the differences between uniform and dedicated caching decrease.

The differences between uniform and dedicated caching depend on the number of gateway nodes. Due to the additional overhead, uniform caching suffers from a larger number of gateway nodes. In contrast, dedicated caching suffers from many gateway nodes, since the probability that cache hits in case of multi-term queries decreases. To quantify this, Figure~\ref{fig:MDK_vs_MDKc_NumOfGN} exemplarily shows the relative results between MDK and MDK$_\mathrm{C}$ and for different number of gateway nodes (we set $c^{ins}=12$ and $c^{del}$=0). The most important result is that for large number of gateway nodes now dedicated caching performs better than uniform caching in terms of number of transferred resources. The explanation is that the bandwidth consumption to forward popular keys to all gateway nodes can no longer be compensated by reduced bandwidth required for processing queries. For dedicated caching the number of transferred resources also increases with larger number of gateway notes, but less pronounced compared to uniform caching. This indicates that the high impact of single-term queries, for which uniform and dedicated caching yield the same results, on the overall performance benefit due to caching.
\\
\\
\textbf{STK$\mathrm{_\mathbf{C}}$ vs MTK$\mathrm{_\mathbf{C}}$.} In our last experiment we compared $\mathrm{STK_C}$ and $\mathrm{MTK_C}$ to quantify the effect of support of multi-term keys in a cache setting. Figure~\ref{fig:trans_STK_MTK} clearly shows difference between uniform and dedicated caching. Since uniform caching reduces the number of transferred resources when processing queries significantly, the additional overhead for resuming multi-term keys in the index has an high impact on the overall performance compared to $\mathrm{STK_C}$. This effect is particularly pronounced in the number of handled resources on back end nodes (see Figure~\ref{fig:load_STK_MTK}) where the index maintenance, including the resuming of multi-term keys, takes places. And since the load of the back end nodes for $\mathrm{STK_C}$ and in case of a large case is almost zero, the difference between are large. For increasing values of $c^{ins}$, the index maintenance overhead for $\mathrm{MTK_C}$ decreases compare to $\mathrm{STK_C}$ since the positive effect of uniform caching on both the number of transferred and handled resources quickly decrease. Compared to uniform caching, for dedicated caching the support of multi-term keys is beneficial in every respect, i.e. regarding the number of transferred resources, see Figure~\ref{fig:trans_STK_MTK}, and the load of both the gateway and back end nodes (Figure~\ref{fig:SDKc_vs_MDKc_scaled} shows the results for dedicated caching with respect to the load of the gateway back end nodes in more detail).
%
\\
\\
As anticipated, the system performance benefits significantly from caching. The more interesting results are: 
(a) Compared to the benefits for the query processing performance, the propagation of updates on the tag data adds only a negligible overhead to the cache maintenance. This is true even for large cache sizes and caching of multi-term keys.
(b) The results uniform and dedicated caching -- or mixed alternatives -- clearly reflects the trade-off between minimizing the traffic in the back end and shifting the overall load from the back end to the gateway nodes, and vice versa. Thus, choice for the caching technique is a design decision particularly depending on the underlying hardware architecture. While similar hardware resources for gateway and back end nodes recommend dedicated caching, a system with powerful gateway nodes and rather low-cost back end nodes will profit from uniform caching.
(c) While systems distributed index storing only single-term keys benefit from caching, in case of dedicated caching the additional support of multi-term keys further boosts the overall system performance. Indexing and caching multi-term keys particularly reduces the number of transferred resources and the load of on the gateway nodes. Only the handling of updates of multi-term keys adds to the load of the back end nodes.

\section{Conclusions}
\label{sec:conclusions}
NoSQL systems are the market's pragmatic answer to meet the need for large-scale distributed storage systems with very high availability.
As their basic data structure, NoSQL systems deploy a key to value map which allows for simple lookups in distributed settings. The hash table like interface inherently limits the efficient evaluation of complex queries. The support of queries apart from the access via the key of data objects, like keyword-based searches, requires additional mechanisms. Divide \& Conquer approaches like \textsc{MapReduce} contact all nodes in the system for each query, potentially leading to an unnecessarily high consumption of resources. Alternatively, various existing NoSQL systems natively support inverted indexes. However, a keyword-based search solely using a single-term inverted index scales poorly in terms of bandwidth consumption in large distributed systems.

We have, therefore, proposed a tagging platform based on a multi-term inverted index where we store adaptively also the inverted lists of popular combinations of terms in the index. Whether a multi-term key is indexed or not depends on its popularity which we derive from the recent query history. We further considered the caching of the most popular single-term and multi-term keys on gateway nodes, i.e., a rather small set of network nodes accepting and handling queries. In our experiments, even with our rather worst-case assumptions and parameter settings, our approaches significantly reduce the overall bandwidth consumption even in the presence of high update rates. The additional storage required to keep the multi-term keys in the index is reasonably small. This increases the capacity of the infrastructure and allows for, e.g., downsizing the deployed resources without sacrificing the performance, thus saving money in terms of installation and operation costs, or conversely, improve the scalability of the existing infrastructure.

\bibliographystyle{abbrv}

\end{document}